\documentclass[]{aa}

\usepackage[varg]{txfonts}
\usepackage{graphicx}
\usepackage{multirow}
\newcommand{\Rs}{$R_\sun{}$}
\usepackage{siunitx}
        \DeclareSIUnit[number-unit-product=\,]\au{au}
        \DeclareSIUnit[number-unit-product=\,]\nT{\nano\tesla}
        \DeclareSIUnit[number-unit-product=\,]\Rs{\textit{R}_\sun{}}
\begin{document}
        \title{Solar-wind predictions for the Parker Solar Probe orbit}
        \subtitle{Near-Sun extrapolations derived from an empirical solar-wind model based on Helios and OMNI observations}
        \author{M.~S.~Venzmer
        \and V.~Bothmer}
        \institute{University of Goettingen, Institute for Astrophysics, Friedrich-Hund-Platz~1, 37077~Göttingen, Germany}
        \date{Received 25 August 2017; accepted 10 November 2017}

        \abstract
        {The Parker Solar Probe (PSP) (formerly Solar Probe Plus) mission will be humanity’s first in situ exploration of the solar corona with closest perihelia at \num{9.86}~solar radii (\si{\Rs}) distance to the Sun. It will help answer hitherto unresolved questions on the heating of the solar corona and the source and acceleration of the solar wind and solar energetic particles. The scope of this study is to model the solar-wind environment for PSP’s unprecedented distances in its prime mission phase during the years \numrange{2018}{2025}. The study is performed within the Coronagraphic German And US SolarProbePlus Survey (CGAUSS)   which is the German contribution to the PSP mission as part of the Wide-field Imager for Solar PRobe (WISPR).}
        {We present an empirical solar-wind model for the inner heliosphere which is derived from OMNI and Helios data. The German-US space probes Helios~1 and Helios~2 flew in the 1970s and observed solar wind in the ecliptic within heliocentric distances of \SIrange{0.29}{0.98}{\au}. The OMNI database consists of multi-spacecraft intercalibrated in situ data obtained near \SI{1}{\au} over more than five solar cycles. The international sunspot number (SSN) and its predictions are used to derive dependencies of the major solar-wind parameters on solar activity and to forecast their properties for the PSP mission.}
        {The frequency distributions for the solar-wind key parameters, magnetic field strength, proton velocity, density, and temperature, are represented by lognormal functions. In addition, we consider the velocity distribution’s bi-componental shape, consisting of a slower and a faster part. Functional relations to solar activity are compiled with use of the OMNI data by correlating and fitting the frequency distributions with the SSN. Further, based on the combined data set from both Helios probes, the parameters’ frequency distributions are fitted with respect to solar distance to obtain power law dependencies. Thus an empirical solar-wind model for the inner heliosphere confined to the ecliptic region is derived, accounting for solar activity and for solar distance through adequate shifts of the lognormal distributions. Finally, the inclusion of SSN predictions and the extrapolation down to PSP’s perihelion region enables us to estimate the solar-wind environment for PSP’s planned trajectory during its mission duration.}
        {The CGAUSS empirical solar-wind model for PSP yields dependencies on solar activity and solar distance for the solar-wind parameters' frequency distributions. The estimated solar-wind median values for PSP’s first perihelion in 2018 at a solar distance of \SI{0.16}{\au} are \SI{87}{\nT}, \SI{340}{\km\per\s}, \SI{214}{\per\cm\cubed} and \SI{503000}{\K}. The estimates for PSP’s first closest perihelion, occurring in 2024 at \SI{0.046}{\au} (\SI{9.86}{\Rs}), are \SI{943}{\nT}, \SI{290}{\km\per\s}, \SI{2951}{\per\cm\cubed}, and \SI{1930000}{\K}. Since the modeled velocity and temperature values below approximately \SI{20}{\Rs} appear overestimated in comparison with existing observations, this suggests that PSP will directly measure solar-wind acceleration and heating processes below \SI{20}{\Rs} as planned.}
        {}
        \keywords{solar wind -- sun: heliosphere -- sun: corona}

        \maketitle

        \section{Introduction}
        From observations of cometary tail fluctuations, \citet{Biermann1951} inferred the presence of a continuous flow of particles from the Sun. With his theoretical solar-wind model, \citet{Parker1958} formulated the existence of the solar wind even before the first satellites measured it in situ in 1959 \citep{Gringauz1960,Neugebauer1966}.
        The idea of a space mission flying through the solar corona dates back to the founding year of NASA in 1958 \citep{McComas2008}. Since then several space missions have measured the solar wind in situ at a wide range of heliocentric distances. In the case of Voyager~1, this was as far away as \SI{140}{\au}\footnote{\url{https://voyager.jpl.nasa.gov/}} in October 2017, having crossed the heliospause into interstellar space at a distance of \SI{121}{\au} \citep{Gurnett2013}.
        Various spacecraft have provided a wealth of solar-wind measurements near Earth’s orbit, with WIND \citep{Lepping1995,Ogilvie1995}, SOHO \citep{Domingo1995}, ACE \citep{Stone1998} and DSCOVR \citep{Burt2012} currently providing observations while orbiting around the L1 point \SI{1.5}{million\,\km} ahead of Earth in the sunward direction. Additional measurements at other solar distances were provided by planetary missions to Venus and Mercury, such as PVO \citep{Colin1980} or MESSENGER \citep{Belcher1991}. Ulysses was the first probe that orbited the Sun out of the ecliptic plane and thus could measure solar wind even at polar latitudes \citep{McComas1998}. The in situ solar-wind measurements closest to the Sun to date were made by the Helios missions. Helios~1, launched in 1974, reached distances of \SI{0.31}{\au}. Helios~2, launched two years later, approached the Sun as close as \SI{0.29}{\au} \citep{Rosenbauer1977}.
        The NASA Parker~Solar~Probe\footnote{\url{http://parkersolarprobe.jhuapl.edu/}} (PSP), formerly Solar~Probe~Plus, six years after its planned launch date in mid 2018, will reach its closest perihelia at a distance of 9.86~solar radii (\Rs), that is, \SI{0.0459}{\au} \citep{Fox2015}. This distance will be achieved through seven Venus gravity assists with orbital periods of 88--168~days. In its prime mission time 2018--2025 PSP provides 24~orbits with perihelia inside \SI{0.25}{\au} \citep{Fox2015}. Even its first perihelion, 93~days after launch in 2018, will take PSP to an unprecedented distance of \SI{0.16}{\au} (\SI{35.7}{\Rs}). In comparison, the ESA Solar Orbiter mission with a planned launch in February 2019 will have its closest perihelia at \SI{0.28}{\au} \citep{Muller2013}.

        The key PSP science objectives are to “trace the flow of energy that heats and accelerates the solar corona and solar wind, determine the structure and dynamics of the plasma and magnetic fields at the sources of the solar wind, and explore mechanisms that accelerate and transport energetic particles” as stated in \citet{Fox2015}. To achieve these goals, PSP has four scientific instruments on board: FIELDS for the measurement of magnetic fields and AC/DC electric fields \citep{Bale2016}, SWEAP for the measurement of flux of electrons, protons and alphas \citep{Kasper2016}, IS\sun{}IS for the measurement of solar energetic particles (SEPs) \citep{McComas2016} and WISPR for the measurement of coronal and inner heliospheric structures \citep{Vourlidas2016}.

        The study presented in this paper is undertaken in the Coronagraphic German And US SolarProbePlus Survey (CGAUSS) project, which is the German contribution to the PSP mission as part of the Wide-field Imager for Solar PRobe (WISPR). WISPR will contribute to the PSP science goals by deriving the three-dimensional structure of the solar corona through which the in situ measurements are made to determine the sources of the solar wind. It will provide density power spectra over a wide range of structures (e.g., streamers, pseudostreamers and equatorial coronal holes) for determining the roles of turbulence, waves, and pressure-balanced structures in the solar wind. It will also measure the physical properties, such as speed and density jumps of SEP-producing shocks and their coronal mass ejection (CME) drivers as they evolve in the corona and inner heliosphere \citep{Vourlidas2016}.
        In order to help optimize the WISPR and PSP preplanning of the science operations, knowledge of the expected solar-wind environment is needed. For this purpose the solar-wind environment is extrapolated down to the closest perihelion of \SI{9.86}{\Rs} distance to the Sun using in situ solar-wind data from the Helios probes and near \SI{1}{\au} data from various satellites compiled in the OMNI solar-wind database.

        Generally, two types of solar wind are observed in the heliosphere -- slow and fast streams \citep{Neugebauer1966,Schwenn1983}. Slow solar wind has typical speeds of \SI{<400}{\km\per\s} and fast solar wind has speeds \SI{>600}{\km\per\s} \citep[p.~144]{Schwenn1990}. Their different compositions and characteristics indicate different sources and generation processes \citep{McGregor2011a}. Fast streams are found to originate from coronal holes as confirmed by Ulysses' out-of-ecliptic measurements \citep{McComas1998}. The source of slow wind, and its eventually different types \citep{Schwenn1983}, is still a subject of controversial discussions because several scenarios are possible to explain its origin from closed magnetic structures in the solar corona, such as intermittent reconnection at the top of helmet streamers and from coronal hole boundaries \citep{Kilpua2016}. The occurrence frequency of these slow and fast streams varies strongly with solar activity and their interactions lead to phenomena such as stream interaction regions which may persist for many solar rotations ("co-rotating" interaction regions) if the coronal source regions are quasi-stationary \citep{Balogh1999}.
        Embedded in the slow and fast solar-wind streams are transient flows of CMEs -- the faster ones driving shock waves ahead \citep{Gosling1974}. Their rate follows the solar activity cycle and varies in near \SI{1}{\au} measurements between only one CME every couple of days during solar cycle minima up to multiple CMEs observed over several days at times of solar maxima, that is, the CME-associated flow share of the solar wind raises from about \SI{5}{\percent} up to about \SI{50}{\percent} \citep{Richardson2012}.

        It is not known which specific solar-wind type or structure PSP will encounter at a given time during its mission, therefore we extrapolate the probability distributions of the major solar-wind parameters from existing solar-wind measurements and take solar cycle dependencies into account.
        As a baseline we describe the solar-wind environment through the key quantities of a magnetized plasma: magnetic field strength, density and temperature. Furthermore, the bulk flow {velocity} is the defining parameter of the two types of solar wind. Solar-wind quantities, like flux densities, mass flux, and plasma beta, can be directly derived from these four parameters. In the analyses, we treat the solar wind as a proton plasma -- the average helium abundance is about \SI{4.5}{\percent} and in slow wind at solar cycle minimum is even less than \SI{2}{\percent} \citep{Feldman1978,Schwenn1983,Kasper2012}.

        Our approach is to obtain analytical representations of the shapes of the solar-wind parameter’s frequency distributions in Sect.~\ref{sec:frequency_distribution}, of their solar activity dependence in Sect.~\ref{sec:solar_activity_variations} and of their solar distance scaling in Sect.~\ref{sec:solar_distance_dependency}. The solar-wind parameters’ frequency distributions and solar activity dependence is derived from near-Earth solar wind and sunspot number (SSN) time series with a duration of almost five solar cycles. Their distance dependency is derived from Helios solar-wind measurements covering more than two thirds of the distance to the Sun and more than half a solar cycle. From a combination of the obtained frequency distributions, SSN dependence functions, and solar distance dependence functions, a general solar-wind model is built in Sect.~\ref{sec:empirical_solar_wind_model}, representing the solar activity and distance behavior. Finally, this empirical model is fed with SSN predictions and extrapolated to PSP's planned orbital positions in Sect.~\ref{sec:model_extrapolation_to_psp_orbit}.

        \section{Frequency distributions of the solar-wind parameters}
        \label{sec:frequency_distribution}
        The solar-wind parameters are highly variable due to short-term variations from structures such as slow and fast wind streams, interaction regions, and CMEs, whose rate and properties depend on the phase of the solar activity cycle. Hence, for deriving characteristic frequency distributions for the solar-wind parameters, measurements over long-term time spans are needed. The abundance of the near-Earth hourly OMNI data set is ideally suited for this purpose, because to date it spans almost five solar cycles.

\begin{figure*}
                \includegraphics[width=18cm]{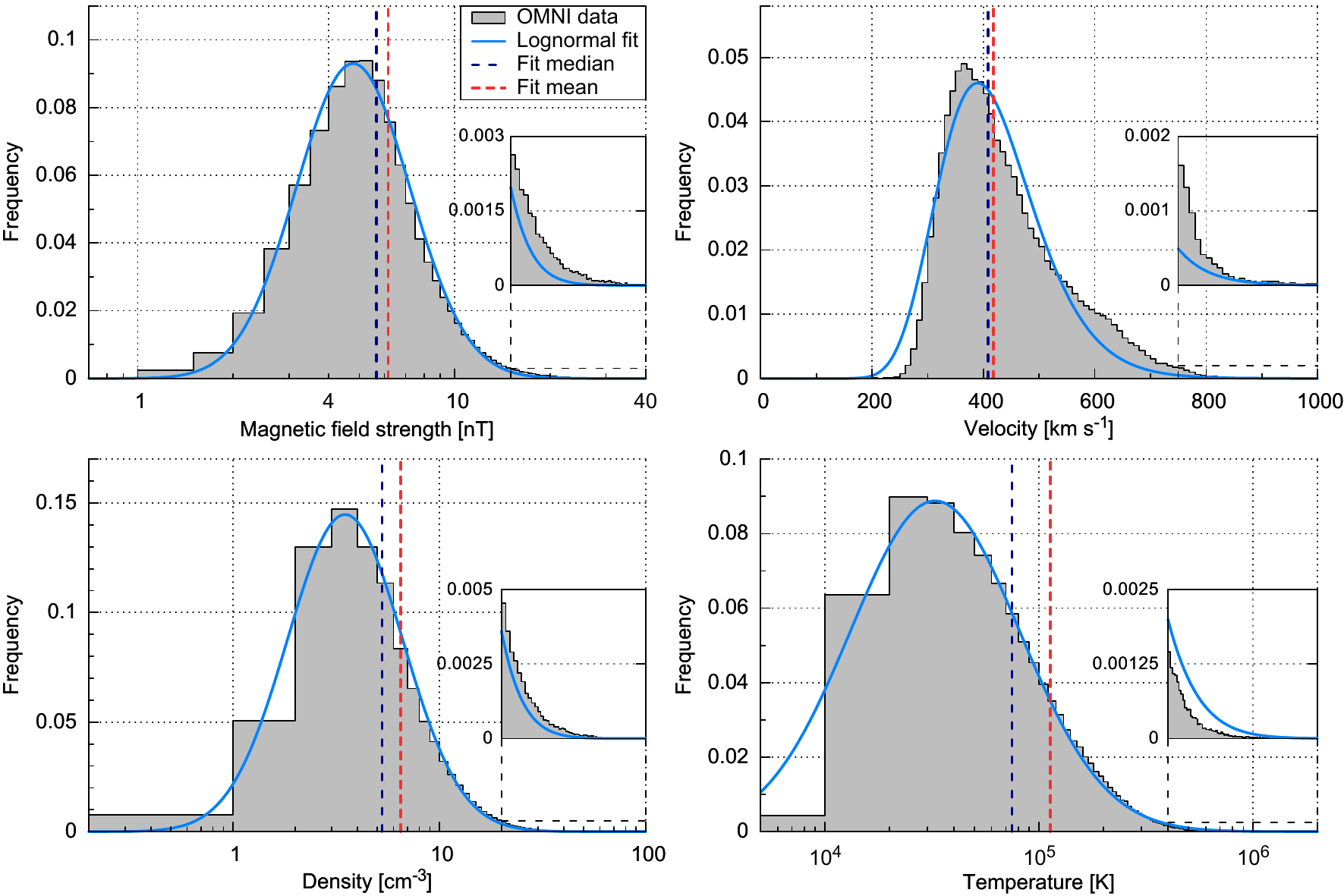}
                \caption{Frequency distributions of the four solar-wind parameters and their lognormal fits derived from the hourly OMNI data set. The histograms have bins of \SI{0.5}{\nT}, \SI{10}{\km\per\s}, \SI{1}{\per\cm\cubed} and \SI{10000}{\K}. The fits' median and mean values are indicated as well. The insets show zoomed-in views of the high-value tails of the distributions.}
                \label{fig:histogram_fits_4_a_zoom_paper_pdfplot}
        \end{figure*}

        The OMNI~2 data set \citep{King2005} combines solar-wind magnetic field and plasma data collected by various satellites since 1963, currently by WIND and by ACE. This intercalibrated multi-spacecraft data is time-shifted to the nose of the Earth’s bow shock. The data is obtained from the OMNIWeb interface\footnote{\url{http://omniweb.gsfc.nasa.gov/}} at NASA's Space Physics Data Facility (SPDF), Goddard Space Flight Center (GSFC).
        In this study the whole hourly data until 31 December 2016 is used, starting from 27 November 1963 (for the temperature from 26 July 1965). The data coverage of the different parameters is in the range \SIrange{67}{74}{\percent},  corresponding to a total duration of 36--40~years.
        We note that a test-comparison of hourly averaged data with higher-time-resolution data for the available shorter time span 1981--2016 did not show significant differences in our results.
        According to the OMNI data precision and maximal parameter ranges we specify bin sizes of \SI{0.5}{\nT} for the magnetic field strength, \SI{10}{\km\per\s} for the velocity, \SI{1}{\per\cm\cubed} for the density and \SI{10000}{\K} for the temperature. The frequency distributions of the solar-wind magnetic field strength, proton velocity, density and temperature are shown in Fig.~\ref{fig:histogram_fits_4_a_zoom_paper_pdfplot}.
        The solar-wind magnetic field strength is in the range \SIrange{0.4}{62}{nT}, the velocity in the range \SIrange{156}{1189}{\km\per\s}, the density in the range \SIrange{0}{117}{\per\cm\cubed}, the temperature in the range \SIrange{3450}{6.63e6}{\K}, and the mean data values are at \SI{6.28}{\nT}, \SI{436}{\km\per\s}, \SI{6.8}{\per\cm\cubed} and \SI{1.05e5}{\K}. These ranges and mean values are as statistically expected from previous analyses of near \SI{1}{\au} solar-wind data (e.g., Table~3.3 in \citet[p.~39]{Bothmer2007}).
        Much higher or lower peak values at \SI{1}{\au} have been observed in extraordinary events, such as the 23~July 2012 CME with a speed of over \SI{2000}{\km\per\s} and a peak field strength of about \SI{100}{\nT} that was observed by STEREO~A \citep{Russell2013}, or the solar-wind disappearance event observed in May 1999 with density values even down to \SI{0.2}{\per\cm\cubed} \citep{Lazarus2000}.

        The frequency distributions of the solar-wind parameters, magnetic field strength, proton density, and temperature, can be
        well approximated by lognormal distributions, whereas the proton velocity’s frequency has a differing shape, as shown in \citet{Veselovsky2010}. We investigate how well all four solar-wind parameters’ frequency distributions can be represented by lognormal functions, which we use in the process of a least squares regression fitting. The lognormal function, 
        \begin{align}
                W(x) &= \frac{1}{\sigma \sqrt{2 \pi} x} \, \exp\left(- \frac{\left(\ln x - \mu\right)^2}{2 \sigma^2}\right),    \label{eq:lognormal_function}
        \end{align}
        depends on the location $\mu$ and the shape parameter $\sigma$. Changes in $\mu$ affect both the horizontal and vertical scaling of the function whereas $\sigma$ influences its shape. The distribution's median $x_\text{med}$ and mean $x_\text{avg}$ (average) positions are easily interpreted and are directly calculated from $\mu$ and $\sigma$:
        \begin{align}
                x_\text{med} &= \exp\left(\mu\right)    &       &\Longleftrightarrow    &       \mu &= \ln\left(x_\text{med}\right)\,,      \label{eq:lognormal_median}\\
                x_\text{avg} &= \exp\left(\mu + \frac{\sigma^2}{2}\right)       &       &\Longleftrightarrow    &       \sigma &= \sqrt{2 \ln\left(\frac{x_\text{avg}}{x_\text{med}}\right)}\,.        \label{eq:lognormal_mean}
        \end{align}
        It is apparent that the mean is always larger than the median. Replacing the variables $\mu$ and $\sigma$ with these relations, the lognormal function~(\ref{eq:lognormal_function}) becomes
        \begin{align}
                W(x) = \frac{1}{2 \sqrt{\pi \ln\left(\frac{x_\text{avg}}{x_\text{med}}\right)} \, x} \, \exp\left(- \frac{\ln^2\left(\frac{x}{x_\text{med}}\right)}{4 \ln\left(\frac{x_\text{avg}}{x_\text{med}}\right)}\right)\,.     \label{eq:single_lognormal_fit_function}
        \end{align}
        The values of $x_\text{med}$ and $x_\text{avg}$ obtained from fitting the individual solar-wind frequency distributions are listed in Table~\ref{tab:lognormal_fit_parameters}.
        \begin{table*}
                \caption{Resulting fit coefficients from the fitting of the lognormal function (\ref{eq:single_lognormal_fit_function}) to the shape of the solar-wind parameters' frequency distributions from near \SI{1}{\au} OMNI hourly data. For the velocity, the fit parameters of the double lognormal function (\ref{eq:double_lognormal_fit_function}) are also listed, as well as the median and mean values of the resulting velocity fit. The numbers in parentheses are the errors on the corresponding last digits of the quoted value. They are calculated from the estimated standard deviations of the fit parameters. For each parameter, the sum of absolute residuals between data and fit (in percentage of the distribution area) is also listed.}
                \label{tab:lognormal_fit_parameters}
                \centering
                \sisetup{table-figures-integer=1, table-figures-decimal=4, table-figures-exponent=0}
                \begin{tabular}{l c
                        S[table-format = 1.3(2), table-space-text-post = a, table-align-text-post = false]
                        S[table-format = 2.3(2), table-space-text-post = a, table-align-text-post = false]
                        @{}c@{}
                        S[table-format = 2.2]
                        }
                        \hline\hline
                        \multicolumn{2}{l}{\multirow{2}{*}{Parameter}}  &\multicolumn{1}{c}{Median}     &\multicolumn{1}{c}{Mean}       &\multicolumn{1}{c}{Balance}    &\multicolumn{1}{c}{SAR}\\
                        \multicolumn{2}{l}{}    &\multicolumn{1}{c}{$x_\text{med}$}     &\multicolumn{1}{c}{$x_\text{avg}$}     &\multicolumn{1}{c}{$c$}        &\multicolumn{1}{c}{[\%]}\\
                        \hline
                        \multicolumn{2}{l}{Magnetic field [\si{nT}]}    &5.661(16)      &6.164(18)      &--     &6.83\\
                        \multicolumn{2}{l}{Velocity [\SI{e2}{\km\per\s}]}       &4.085(19)      &4.183(20)      &--     &18.69\\
                        \multicolumn{2}{l}{Density [\si{\per\cm\cubed}]}        &5.276(24)      &6.484(34)      &--     &6.48\\
                        \multicolumn{2}{l}{Temperature [\SI{e4}{\K}]}   &7.470(17)      &11.301(32)     &--     &5.78\\
                        \hline
                        \multirow{2}{*}{Velocity}       &\multicolumn{1}{c}{$W_1$}      &4.89(14)       &5.00(14)       &\multirow{2}{*}{0.504(62)}     &\multicolumn{1}{c}{--}\\
                        \multirow{2}{*}{[\SI{e2}{\km\per\s}]~}  &\multicolumn{1}{c}{$W_2$}      &3.68(20)       &3.72(20)       &       &\multicolumn{1}{c}{--}\\
                        \cline{2-6}
                                &\multicolumn{1}{c}{$W_\text{II}$}      &4.16(14)\tablefootmark{a}      &4.42(14)\tablefootmark{a}      &\multicolumn{1}{c}{--} &4.20\\
                        \hline
                \end{tabular}
                \tablefoot{
                        \tablefoottext{a}{Error estimates derived from the individual fit part errors.}
                }
        \end{table*}

\begin{figure}
                \resizebox{\hsize}{!}{\includegraphics{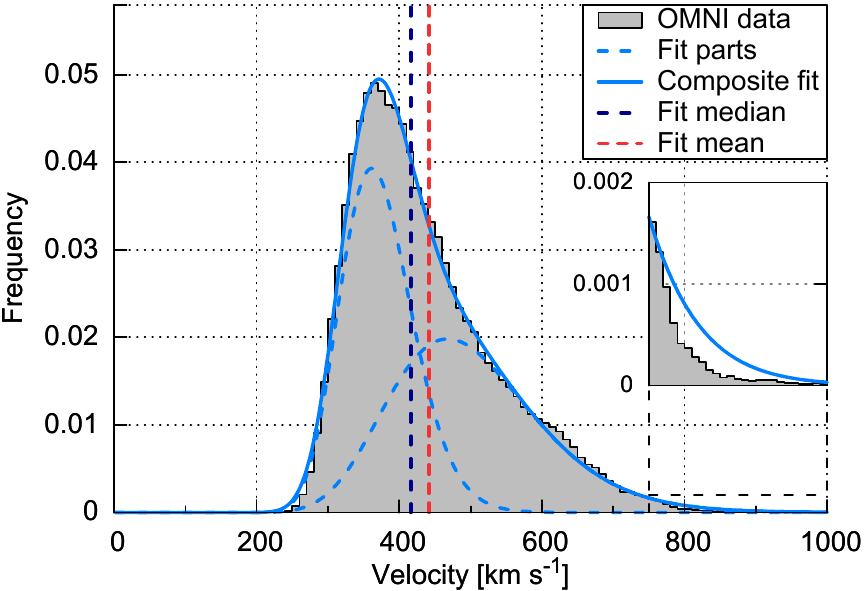}}
                \caption{Velocity frequency distribution (same as in Fig.~\ref{fig:histogram_fits_4_a_zoom_paper_pdfplot}) and its compositional lognormal fit. The fit's median and mean values and its two fit parts are indicated as well. The inset is a zoomed-in view of the high value tail of the distribution.}
                \label{fig:histogram_fits_V_a_zoom_dbl_paper_pdfplot}
        \end{figure}

        From visual inspection, the resulting fit curves describe the shape of the magnetic field strength, the density and the temperature distributions well, as can be seen in Fig.~\ref{fig:histogram_fits_4_a_zoom_paper_pdfplot}.
        However, for the velocity, the fit function appears not to be as good in describing the measured distribution’s more complex shape around its peak and in the higher velocity range. This also can be inferred from the sum of absolute residuals (SAR) between data and fit, listed in Table~\ref{tab:lognormal_fit_parameters} as a percentage of the distribution area, being almost three times larger than those from the other parameters.
        In order to find a better fit result for the velocity distribution, we assume that the velocity distribution can be made up of at least two overlapping branches \citep{McGregor2011b}. Therefore a compositional approach  is chosen by combining two lognormal functions (\ref{eq:single_lognormal_fit_function}), involving more fit variables:
        \begin{align}
                W_\text{II}(x) &= c \cdot W_1(x) + (1 -c) \cdot W_2(x)\,.       \label{eq:double_lognormal_fit_function}
        \end{align}
        The balancing parameter $c$ ensures that the resulting function remains normalized as it represents a probability distribution.
        The fitting of $W_\text{II}(x)$ to the velocity's frequency distribution yields the values of the now five fit parameters ($c$, $x_\text{med,1}$, $x_\text{avg,1}$, $x_\text{med,2}$ and $x_\text{avg,2}$) as listed in Table~\ref{tab:lognormal_fit_parameters} together with the median and mean values of the composed distribution, which can be derived by solving
        \begin{align}
                \int W_\text{II}(x)\,\text{d}x = 0      &       &\text{and}     &       &\int x\,W_\text{II}(x)\,\text{d}x = 0        \,.
        \end{align}
        This more complex fit function is more accurate in describing the velocity's frequency distribution as shown in Fig.~\ref{fig:histogram_fits_V_a_zoom_dbl_paper_pdfplot}.
        Thus in the following Sections we keep the double lognormal ansatz for all velocity frequency fits.

        For the bulk of the solar wind these static lognormal functions describe the parameters' distributions well. The abnormally high parameter values in the distribution functions can be attributed to shock/CME events in agreement with the results of the OMNI solar-wind investigations by \citet{Richardson2012}. The simple lognormal fit functions underestimate the frequencies in their high-value tails, except for the temperature’s tail which is overestimated, as seen in the insets of Fig.~\ref{fig:histogram_fits_4_a_zoom_paper_pdfplot}. This appears to be because CMEs do not come with abnormally high temperatures, but rather with temperatures lower than those of the average solar wind \citep{Forsyth2006}. The velocity's compositional lognormal fit only slightly overestimates its tail as seen in the inset of Fig.~\ref{fig:histogram_fits_V_a_zoom_dbl_paper_pdfplot}.
        The slow and fast part contribute almost equally ($c \approx 0.5$) to the long-term velocity distribution function.

        \section{Solar activity dependence of the solar-wind frequency distributions}
        \label{sec:solar_activity_variations}
        In the next step we investigate how the long-term solar-wind distribution functions presented in the previous section depend on general solar activity. Therefore we examine their correlation with the SSN, being a commonly used long-term solar activity index, and determine the time lags with the highest correlation coefficients.

        For the correlations we fit lognormal functions to the frequency distributions as in Sect.~\ref{sec:frequency_distribution}, but implement linear relations to the yearly SSN, allowing shifting of the distribution functions with SSN. For the velocity the approach is different insofar as its two components are kept fixed and instead their balance is modified with the changing SSN. Thus we obtain solar-activity-dependent models for the frequency distributions of all four solar-wind parameters.

        The international sunspot number (\citeyear{sidc}) is provided by the online catalog\footnote{\url{http://www.sidc.be/silso/}} at the World Data Center -- Sunspot Index and Long-term Solar Observations (WDC-SILSO), Solar Influences Data Analysis Center (SIDC), Royal Observatory of Belgium (ROB).

\begin{figure}
                \resizebox{\hsize}{!}{\includegraphics{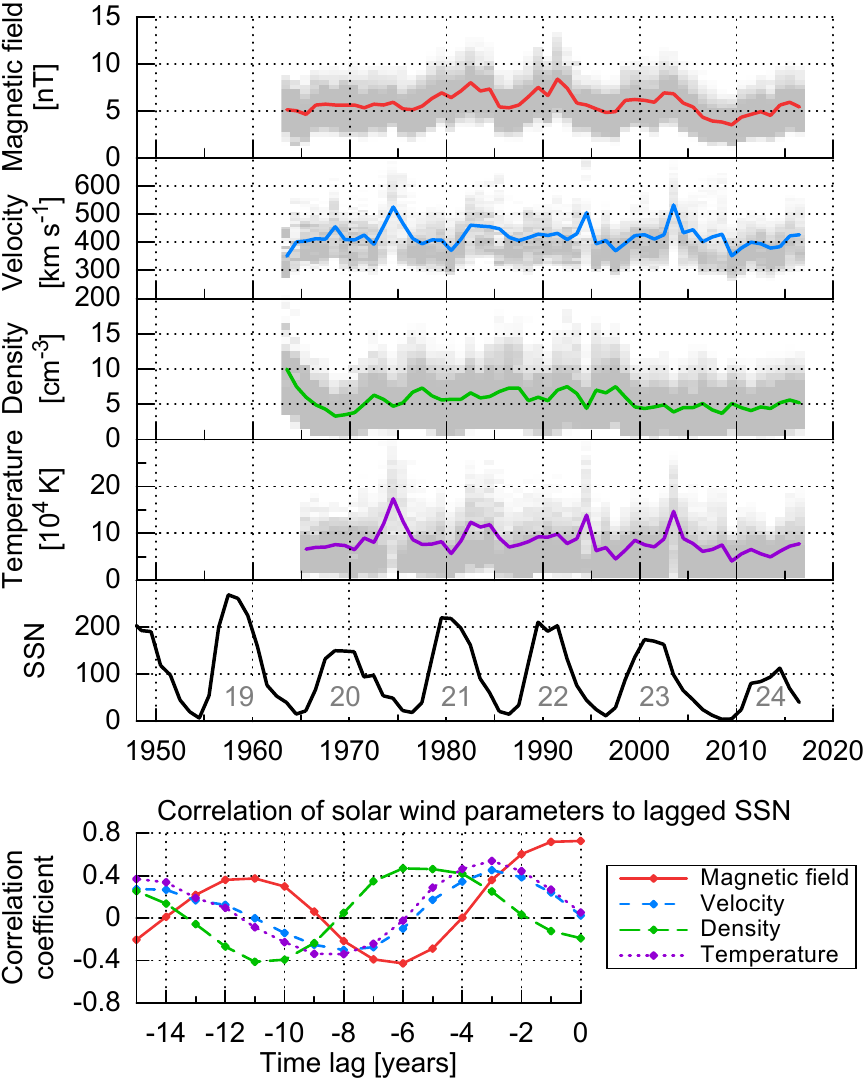}}
                \caption{Solar-wind parameter yearly frequencies (gray shading) with yearly medians (lines) derived from OMNI data and the yearly SSN from the \citet{sidc} with solar cycle number (top). Their correlation coefficients with the yearly SSN are calculated for time lags back to -15 years (bottom).}
                \label{fig:OMNI_yearly_ssn_correlation_c_plot}
        \end{figure}

        Yearly medians of the solar-wind parameters and the yearly SSN together with the solar cycle number are shown in the upper part of Fig.~\ref{fig:OMNI_yearly_ssn_correlation_c_plot}. The reason for correlating the SSN to the solar-wind median values is because the position of a lognormal function is defined by its median. The data are averaged to yearly values to avoid seasonal effects during the Earth’s orbit around the Sun caused by its variations in solar latitude and distance.
        The solar-wind velocity, density, and temperature depend on the state of the solar cycle \citep{Schwenn1983}.
        For instance the fast solar wind occurs at times when polar coronal holes extend to lower latitudes, a typical feature of the declining phase of the solar cycle as pointed out by \citet[p.~75, Fig.~3.52]{Bothmer2007}. Therefore the solar-wind velocity, density, and temperature maxima exhibit time lags relative to the SSN maxima.

        The correlation coefficients of the solar wind parameters with the yearly SSN shown in the bottom part of Fig.~\ref{fig:OMNI_yearly_ssn_correlation_c_plot} are calculated for time lags back to \num{-15}~years to cover a time span longer than a solar cycle. As expected, the amplitudes of the variations in the correlations of all parameters decline with increasing time lag and show a period of about 11~years. The highest correlation coefficient of 0.728 to the SSN is found for the magnetic field strength; it has no time lag. This finding is anticipated because the SSN is found to be directly proportional to the evolution of the photospheric magnetic flux \citep{Smith2003}.
        Velocity and temperature show time lags of 3~years with peak correlation coefficients of 0.453 and 0.540. The density with a correlation coefficient of 0.468 has a time lag of 6~years, which is in agreement with the density anticorrelation to the SSN reported by \citet{Bougeret1984}.

        Next we create solar-activity-dependent analytical representations of the solar wind frequency distributions. This is achieved by shifting the median positions of the lognormal distributions as a linear function of the SSN. To enable these shifts, we add a linear SSN dependency to the median,
        \begin{align}
                x_\text{med}(ssn) &= a_\text{med} \cdot ssn + b_\text{med}\,,   \label{eq:median_with_ssn}
        \end{align}
        using a factor to the SSN $a_\text{med}$ with a baseline $b_\text{med}$. We relate the mean with a scaling factor to the median to transfer its SSN dependency:
        \begin{align}
                x_\text{avg}(ssn) &= \left(1 + a_\text{avg}\right) \cdot x_\text{med}(ssn)\,.    \label{eq:mean_with_ssn}
        \end{align}
        These relations, substituted into the lognormal function (\ref{eq:single_lognormal_fit_function}), lead to a new SSN-dependent function $W'(x,ssn)$. This function is then fitted to the yearly data, using the yearly SSN as input parameter. The SSN is offset with the individual time lags determined before for each parameter, to benefit from the higher correlation. The values of the three resulting fit coefficients ($a_\text{med}, b_\text{med}$ and $a_\text{avg}$) are presented in Table~\ref{tab:ssn_fit_parameters}.
        \begin{table*}
                \caption{Resulting fit coefficients from the OMNI data, based on the linear SSN dependencies (\ref{eq:median_with_ssn}) and (\ref{eq:mean_with_ssn}). For the velocity the fit parameters from the double lognormal fit (\ref{eq:double_lognormal_fit_function}) and their balancing function (\ref{eq:balance_with_ssn}) are given. The numbers in parentheses are the errors on the corresponding last digits of the quoted value. They are calculated from the estimated standard deviations of the fit parameters. The listed SSN time lags are used for the fits.}
                \label{tab:ssn_fit_parameters}
                \centering
                \begin{tabular}{l@{} c@{}
                        S[table-format = 1.3(2)e+1]
                        S[table-format = 1.4(2)]
                        S[table-format = 1.3(2)e+1]
                        S[table-format = +1.3(2)e+1]
                        c c
                        }
                        \hline\hline
                        \multicolumn{2}{l}{\multirow{2}{*}{Parameter}}  &\multicolumn{2}{c}{Median}     &\multicolumn{1}{c}{Mean}       &\multicolumn{2}{c}{Balance}    &\multicolumn{1}{c}{SSN lag}\\
                        \cline{3-4}\cline{6-7}
                        \multicolumn{2}{l}{}    &\multicolumn{1}{c}{SSN factor $a_\text{med}$} &\multicolumn{1}{c}{Baseline $b_\text{med}$}    &\multicolumn{1}{c}{Scaling factor $a_\text{avg}$}  &\multicolumn{1}{c}{SSN factor $c_a$}   &\multicolumn{1}{c}{Baseline $c_b$}  &\multicolumn{1}{c}{[years]}\\
                        \hline
                        \multicolumn{2}{l}{Magnetic field [\si{nT}]}    &1.309(19)e-2   &4.285(17)      &8.786(78)e-2   &\multicolumn{1}{c}{--} &--     &0\\
                        \multicolumn{2}{l}{Density [\si{\per\cm\cubed}]}        &3.81(25)e-3    &4.495(26)      &3.050(27)e-1   &\multicolumn{1}{c}{--} &--     &6\\
                        \multicolumn{2}{l}{Temperature [\SI{e4}{\K}]}   &1.974(26)e-2   &5.729(19)      &6.541(28)e-1   &\multicolumn{1}{c}{--} &--     &3\\
                        \hline
                        \multicolumn{1}{l}{Velocity}    &\multicolumn{1}{c}{$W'_1$}     &\multicolumn{1}{c}{--} &3.633(12)      &1.008(37)e-2   &-1.799(95)e-3      &\multirow{2}{*}{0.638(32)}     &\multirow{2}{*}{3}\\
                        \multicolumn{1}{l}{[\SI{e2}{\km\per\s}]}        &\multicolumn{1}{c}{$W'_2$}     &\multicolumn{1}{c}{--} &4.831(81)      &2.31(20)e-2    &-1.799(95)e-3       &       &\\
                        \hline
                \end{tabular}
        \end{table*}

\begin{figure*}
                \includegraphics[width=18cm]{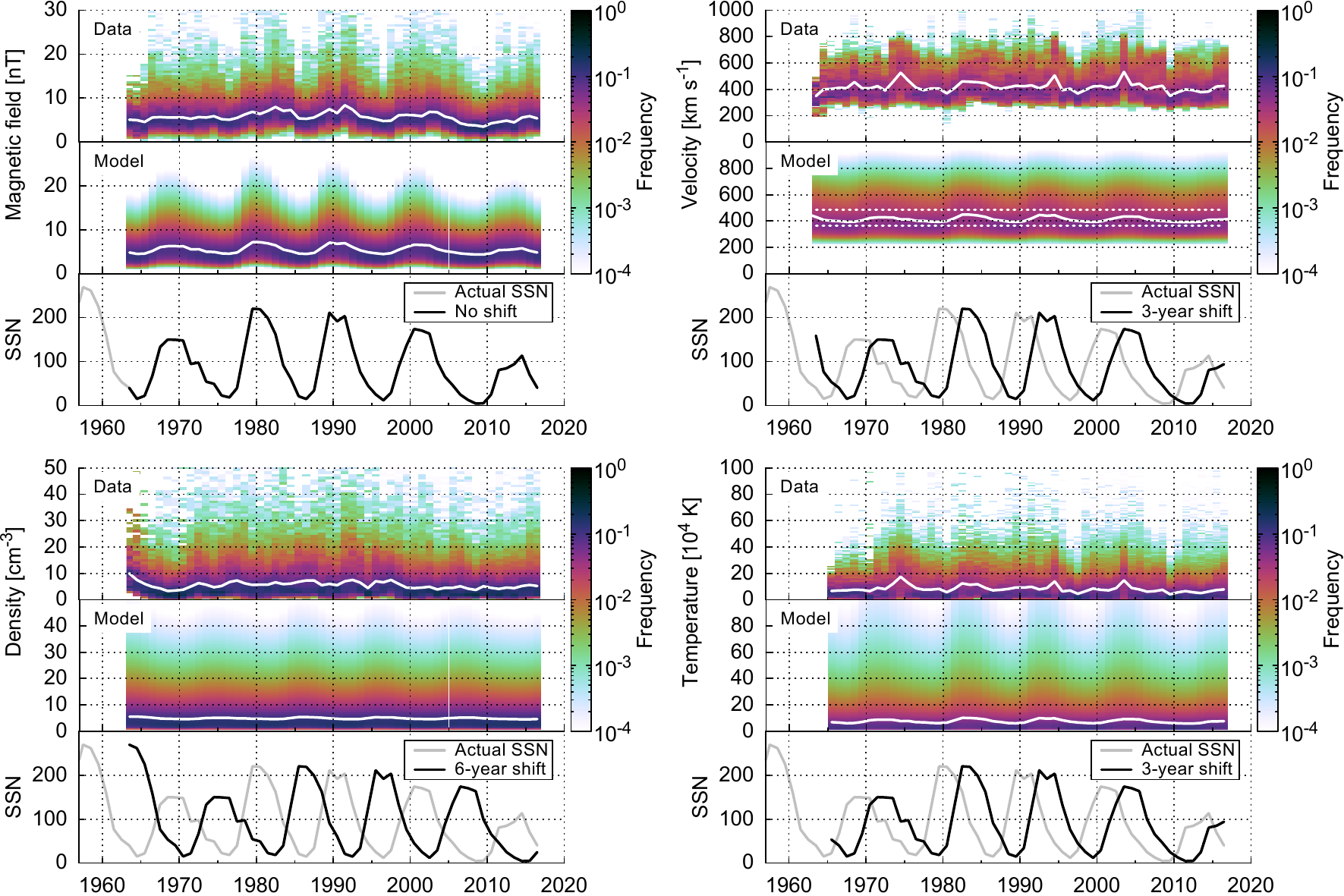}
                \caption{Solar wind parameter yearly data frequencies and lognormal fit models, both with their median values (white lines) over the OMNI time period 1963--2016. The corresponding yearly SSN and the shifted SSN for the models are indicated by gray and black lines. The velocity median is derived from the SSN-weighted constant lognormal parts (dotted lines).}
                \label{fig:OMNI_yearly_BVdblNTSSN_fit_e_plot}
        \end{figure*}

\begin{figure}
                \resizebox{\hsize}{!}{\includegraphics{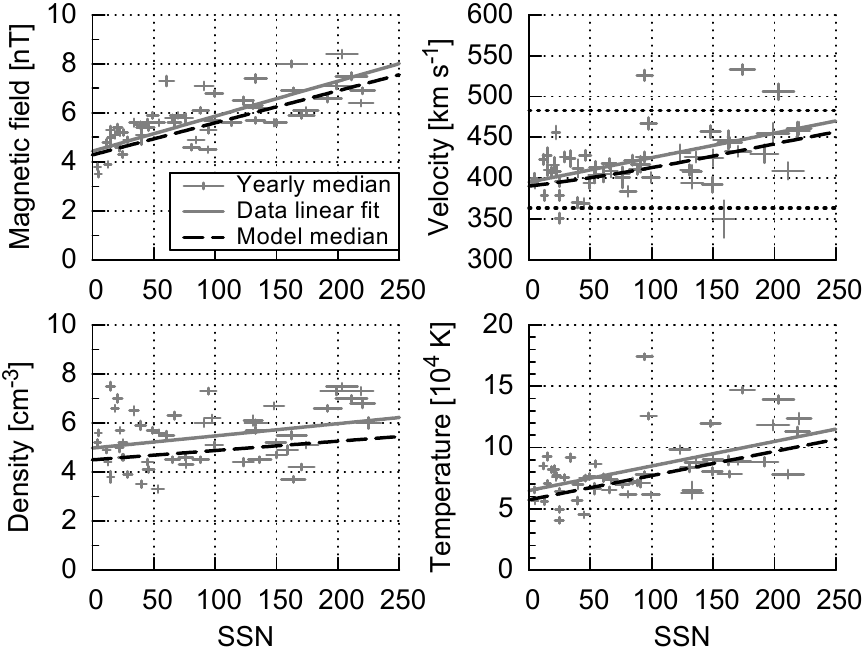}}
                \caption{Solar-wind parameter medians with respect to the lagged SSN. The yearly data medians (+) with their weighted linear fit (solid lines) are obtained from OMNI data. The error bars denote the SSN standard deviation and the relative weight from the yearly data coverage. The SSN-dependent median (dashed lines) is derived from the lognormal model fit. For the velocity the median is derived from the SSN-weighting (\ref{eq:balance_with_ssn}) of the slow and fast model parts (dotted lines), whose magnitudes are SSN independent.}
                \label{fig:OMNI_yearly_BVNTvsSSN_a}
        \end{figure}

        Naturally, the fit models match with the general data trends, as can be seen from Fig.~\ref{fig:OMNI_yearly_BVdblNTSSN_fit_e_plot}, though single year variations are not replicated by the model (e.g., the high velocity and temperature values in 1974, 1994, and 2003).
        The comparison of this model with the yearly data median values with respect to the lagged SSN shows that the medians obtained from the modeling have a similar slope, as shown in Fig.~\ref{fig:OMNI_yearly_BVNTvsSSN_a}.

\begin{figure}
                \resizebox{\hsize}{!}{\includegraphics{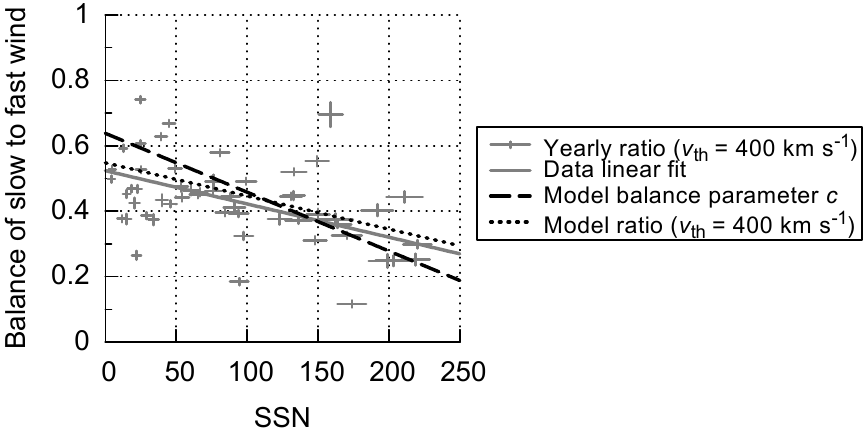}}
                \caption{Ratio of slow to fast solar wind for a SSN lagged by 3~years. The yearly ratios (+) and their weighted linear fit (solid line) are obtained from OMNI data with a threshold velocity of $v_\text{th} = \SI{400}{\km\per\s}$. The error bars denote the SSN standard deviation and the relative weight from the yearly data coverage. The model's balance parameter (\ref{eq:balance_with_ssn}) and derived ratio (same threshold) are plotted as dashed and dotted lines.}
                \label{fig:Vdbl_SSN_ratio_f_plot}
        \end{figure}

        Again, the solar-wind velocity needs a special treatment because of the application of the double lognormal distribution (\ref{eq:double_lognormal_fit_function}). Since it is well known that slow and fast solar-wind stream occurrence rates follow the solar cycle, we keep the two velocity components' positions SSN-independent ($x_\text{med} =  b_\text{med}$) and vary instead their balance with the SSN:
        \begin{align}
                c(ssn) &= c_a \cdot ssn + c_b\,.        \label{eq:balance_with_ssn}
        \end{align}
        The fit result (see Table~\ref{tab:ssn_fit_parameters}) yields a model in which three years after solar cycle minimum (SSN of zero) the contribution of slow solar wind to the overall solar wind distribution reaches a maximum value (about \SI{64}{\percent}) and decreases with increasing SSN as shown in Fig.~\ref{fig:Vdbl_SSN_ratio_f_plot}.

        To investigate the amount of slow and fast wind contributions depending on solar activity, we apply the commonly used constant velocity threshold of $v_\text{th} = \SI{400}{\km\per\s}$ \citep[p.~144]{Schwenn1990}. The linear fit to the yearly data ratio and the derived model ratio show a good agreement (see Fig.~\ref{fig:Vdbl_SSN_ratio_f_plot}). The to-some-degree steeper balance parameter of the double fit function used in this model cannot be compared directly with specific velocity thresholds between slow and fast solar wind. However, it appears to be a more realistic approach than just taking a specific velocity threshold for the slow and fast wind, in agreement with the overlapping nature of the velocity flows reported by \citet{McGregor2011b}.

        \section{Solar distance dependency}
        \label{sec:solar_distance_dependency}
        In order to derive heliocentric distance relationships of the bulk solar wind distribution functions, we apply and fit power law dependencies to the Helios data. We then examine how the fits may be extrapolated towards the Sun and in particular in to the PSP orbit. We use the fitting methods of Sect.~\ref{sec:frequency_distribution} for the distance-binned combined data from both Helios probes. Helios’ highly elliptical orbits in the ecliptic covered a solar distance range of \SIrange{0.31}{0.98}{\au} in case of Helios~1 and \SIrange{0.29}{0.98}{\au} in case of Helios~2. Launched during solar cycle minimum, the data of both probes cover the rise to the maximum of cycle~21, covering $\sim$6.5~years at varying distances to the Sun.

        We investigate hourly averages of the Helios data in the same way as with the OMNI data. The Helios~1 merged hourly data from the magnetometer and plasma instruments \citep{Rosenbauer1977} include $\sim$12.5~orbits for the time range 10~December 1974 to 14~June 1981, and those for Helios~2 include $\sim$8~orbits for the time span 1~January 1976 to 4~March 1980. The data are retrieved from the Coordinated~Data~Analysis~Web (CDAWeb) interface at NASA's GSFC/SPDF\footnote{\url{http://spdf.gsfc.nasa.gov/}}.

        The Helios~1 magnetometer data coverage for this data set is about \SI{43}{\%} (i.e., 2.8~years), and that of Helios~2 amounts to \SI{54}{\%} (i.e., 2.3~years). The plasma data coverage is \SI{76}{\%} (i.e., 5.0~years) in case of Helios~1 and \SI{92}{\%} (i.e., 3.9~years) in case of Helios~2. 
        Thus, using this data, we point out that its time coverage is unequally distributed over the solar cycle. Considering the data gap distributions, the amount of data during solar cycle minimum up to mid~1977, that is, the transition from minimum to maximum, covers about \SI{68}{\percent} of this period whereas during maximum of cycle~21 data are available only \SI{38}{\percent} of the time. This Helios data bias towards solar minimum is one reason why in this study the Helios solar wind data are not used to derive long-term frequency distributions and solar-cycle dependencies for the key solar wind parameters.

\begin{figure*}
                \includegraphics[width=18cm]{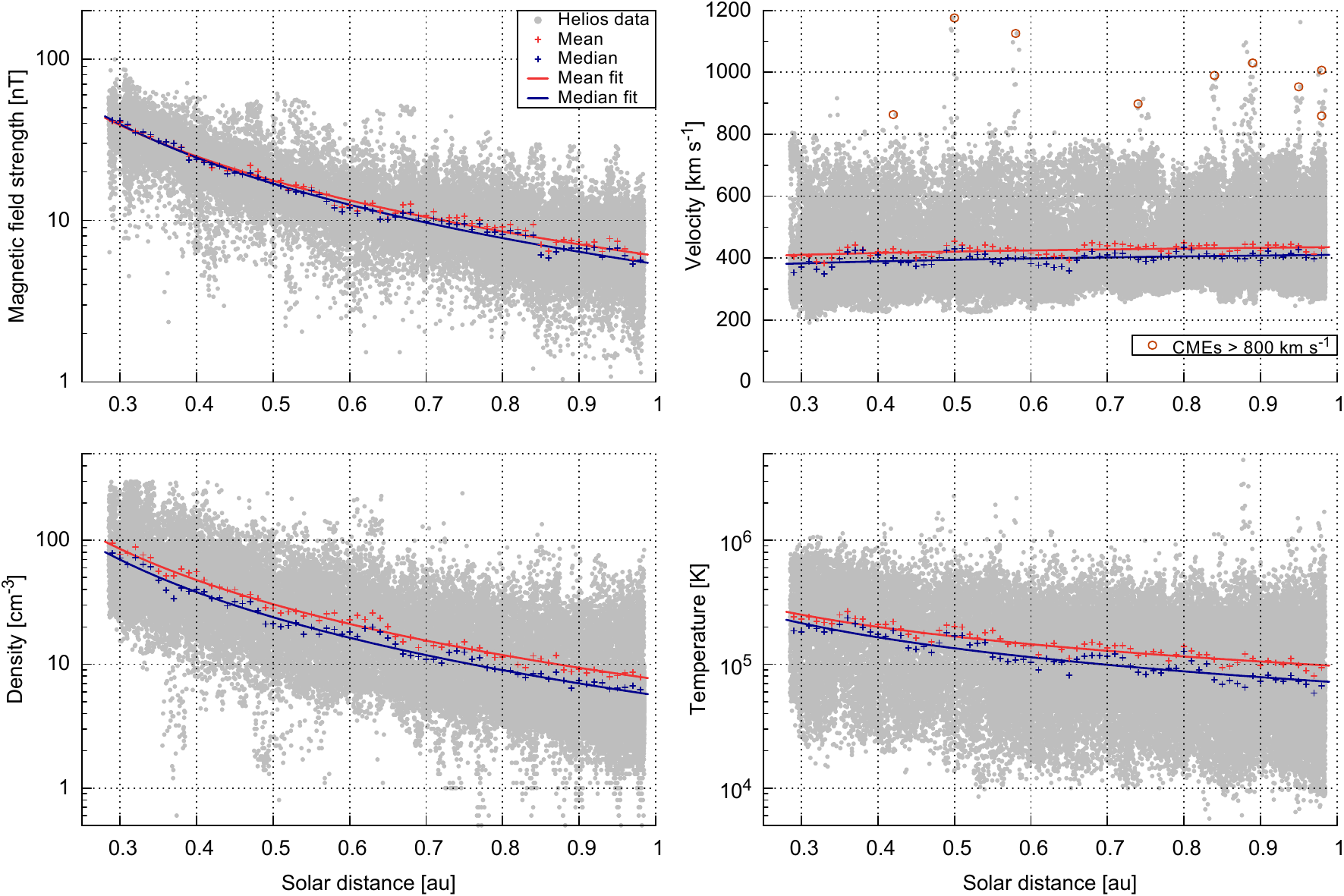}
                \caption{Helios hourly data plots of the four solar wind parameters over solar distance. The mean and median per \SI{0.01}{\au} data bin and their fit curves are plotted as well. The Helios data has a native distance resolution of \SI{0.01}{\au}, thus, to make the distribution visible in these plots, we added a random distance value of up to \SI{+-0.005}{\au}. The high velocity data points above \SI{800}{\km\per\s} (circled red) are identified as CME events \citep[e.g.,][]{Sheeley1985,Bothmer1996,Bothmer1998}.}
                \label{fig:radial_fit_4_thesis_light_b_skip}
        \end{figure*}

        The radial dependencies of the key solar-wind parameters over the distance range \SIrange{0.29}{0.98}{\au} measured by both Helios probes are plotted in Fig.~\ref{fig:radial_fit_4_thesis_light_b_skip}, together with their median and mean values for different solar distances, calculated for the minimal distance resolution \SI{0.01}{\au} of the data set.
        Assuming a radial solar-wind outflow, it is expected that the distance dependence of the solar-wind parameters over the Helios data range \SIrange{0.29}{0.98}{\au} can be described through power law scaling. Therefore we use the power law function,
        \begin{align}
                x(r) = d\cdot r^e,       \label{eq:power_function}
        \end{align}
        for the regression fit of the median and mean, with $r$ being the solar distance in astronomical units, $d$ the magnitude at \SI{1}{\au} and $e$ the exponent. The fits are weighted through the different data counts per bin.
        The obtained coefficients for the median and mean power law fits ($d_\text{med}$, $e_\text{med}$, $d_\text{avg}$ and $e_\text{avg}$) are listed in Table~\ref{tab:mean_median_fit_parameter} and their corresponding curves are shown in Fig.~\ref{fig:radial_fit_4_thesis_light_b_skip}.
        \begin{table*}
                \caption{Fit coefficients for the median and mean solar distance dependencies (\ref{eq:power_function}) of the four solar wind parameters derived from the combined Helios~1 and 2 data. The numbers in parentheses are the errors on the corresponding last digits of the quoted value. They are calculated from the estimated standard deviations of the fit parameters. The crossing distances indicate where the median and mean fits intersect each other. The yearly variation is the weighted standard deviation derived from the yearly fit exponents seen in Fig.~\ref{fig:yearly_gradients_c}.}
                \label{tab:mean_median_fit_parameter}
                \centering
                \begin{tabular}{l
                S[table-format = 1.3(2)]
                S[table-format = +1.3(2)]
                @{}c
                S[table-format = 1.3(2)]
                S[table-format = +1.3(2)]
                S[table-format = 1.1(2)e1]
                S[table-format = 1.3]}
                        \hline\hline
                        \multirow{2}{*}{Parameter}      &\multicolumn{2}{c}{Median}     &       &\multicolumn{2}{c}{Mean}       &\multicolumn{1}{c}{Crossing distance}       &\multicolumn{1}{c}{Yearly variation}\\
                        \cline{2-3}     \cline{5-6}
                                &\multicolumn{1}{c}{$d_\text{med}$}     &\multicolumn{1}{c}{$e_\text{med}$}     &       &\multicolumn{1}{c}{$d_\text{avg}$}     &\multicolumn{1}{c}{$e_\text{avg}$}     &\multicolumn{1}{c}{[\si{\au}]} &\multicolumn{1}{c}{$\Delta e$}\\
                        \hline
                        Magnetic field [\si{nT}]        &5.377(92)      &-1.655(17)     &       &6.05(10)       &-1.546(18)     &0.339(11)      &0.11\\
                        Velocity [\SI{e2}{\km\per\s}]   &4.107(28)      &0.058(13)      &       &4.356(24)      &0.049(10)      &0.7(83)e3      &0.012\\
                        Density [\si{\per\cm\cubed}]    &5.61(27)       &-2.093(46)     &       &7.57(30)       &-2.010(38)     &0.027(73)      &0.072\\
                        Temperature [\SI{e4}{\K}]       &7.14(23)       &-0.913(39)     &       &9.67(21)       &-0.792(28)     &0.082(85)      &0.050\\
                        \hline
                \end{tabular}
        \end{table*}

        Our derived exponents agree with those found in existing studies from the Helios observations: \citet{Mariani1978} derived the exponents for the magnetic field strength separately for the fast and slow solar wind as $B_\text{fast} \propto r^{-1.54}$ and $B_\text{slow} \propto r^{-1.61}$, ours is $B_\text{avg} \propto r^{-1.55}$.
        The velocity exponent $v_\text{avg} \propto r^{0.049}$ matches with the values found by \citet{Schwenn1983,Schwenn1990}, who derived the distance dependencies for both Helios spacecraft separately as $v_\text{H1} \propto r^{0.083}$ and $v_\text{H2} \propto r^{0.036}$. The calculated density exponent $n_\text{avg} \propto r^{-2.01}$ agrees well with the Helios plasma density model derived by \citet{Bougeret1984} yielding $n \propto r^{-2.10}$.
        The temperature exponent $T_\text{avg} \propto r^{-0.79}$ is similar to those in the studies by \citet{Hellinger2011,Hellinger2013}, who also derived the exponents separately for the fast and the slow solar wind: $T_\text{fast} \propto r^{-0.74}$ and $T_\text{slow} \propto r^{-0.58}$.

        The mean and median velocity fit exponents acquired from the Helios data are very similar, which indicates that they can be kept identical so that the basic shape of the frequency distribution does not change with distance. Conversely, the mean and median fits for the magnetic field strength cross each other at \SI{0.339}{\au} (see Table~\ref{tab:mean_median_fit_parameter}) and the mean becomes slightly lower than the median at smaller distances. Thus, below that distance the frequency distribution can no longer be well described by a lognormal function, because the mean of a lognormal function has to be larger than its median (as pointed out in Sect.~\ref{sec:frequency_distribution}), that is, the location of the crossing indicates that the parameter's distribution is no longer of a lognormal shape thereafter. The fits for the proton temperature show a similar behavior, having an extrapolated intersection at \SI{0.082}{\au}. Therefore the extrapolation of the magnetic field and temperature distribution frequencies to the PSP orbit by applying lognormal functions is limited. The crossing points limit the regions where the distribution's shapes can still be considered lognormal.

        In order to still fit and extrapolate lognormal functions with the data, we assume that the shapes can be considered lognormal at all distances. For the frequency distribution fit function to be discussed in the following paragraph, we reduce the fit exponents $e_\text{med}$ and $e_\text{avg}$ to only one. We note that this simplification leads to slightly larger modeling errors, especially in case of the magnetic field strength.

\begin{figure*}
                \includegraphics[width=18cm]{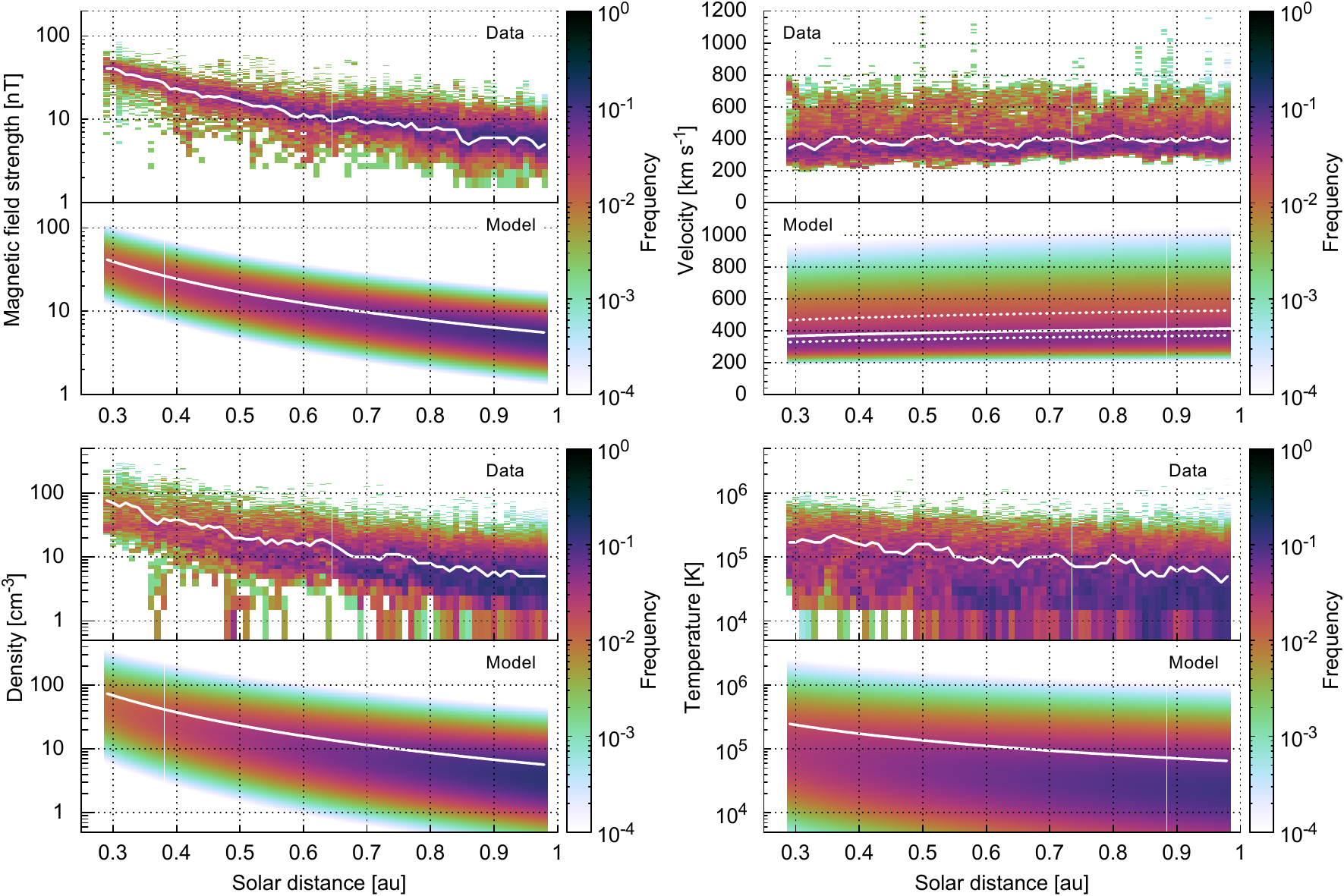}
                \caption{Frequency distributions of the four solar wind parameters with respect to solar distance. Plotted are the binned Helios data and the power law lognormal fit models with their median values (white lines). The double lognormal model is used for the velocity, its slow and fast parts are indicated by dotted lines.}
                \label{fig:mixed_fit_fixed_4_paper_f_plot}
        \end{figure*}

        Next we retrieve the frequency distributions of the four solar wind parameters in solar distance bins of \SI{0.01}{\au}, choosing the same resolution as for the OMNI data analyzed in Sect.~\ref{sec:frequency_distribution} -- the distributions and their median values are plotted in Fig.~\ref{fig:mixed_fit_fixed_4_paper_f_plot}.
        For simplification, as mentioned before, we treat the exponents of the median and mean fit functions as being identical, using one fit parameter for both. Implementing the power law distance dependency~(\ref{eq:power_function}) into the lognormal function (\ref{eq:single_lognormal_fit_function}), we get the fit parameters $d'_\text{med}$, $d'_\text{avg}$ and the common exponent $e'$. Again, we use the double lognormal function~(\ref{eq:double_lognormal_fit_function}) for the velocity distribution fit -- resulting in $W''_\text{II}(x,r)$. The additional fit parameters are the balancing parameter $c'$ and for the second lognormal part $d'_\text{med,2}$ and $d'_\text{avg,2}$. The resulting fit coefficients for the four solar wind parameters are presented in Table~\ref{tab:extrapolation_model_fit_parameters}.
        \begin{table*}
                \caption{Fit coefficients for the distance-dependent single lognormal function, based on Equation (\ref{eq:single_lognormal_fit_function}) combined with (\ref{eq:power_function}) from the combined Helios data. Regarding the velocity, the double lognormal function (\ref{eq:double_lognormal_fit_function}) is used instead. The numbers in parentheses are the errors on the corresponding last digits of the quoted value. They are calculated from the estimated standard deviations of the fit parameters. The seasonal variations are calculated from Earth's orbital solar distance variation and the derived exponents.}
                \label{tab:extrapolation_model_fit_parameters}
                \centering
                \sisetup{table-figures-integer=1, table-figures-decimal=3, table-figures-exponent=0}
                \begin{tabular}{l c
                S[table-format = 1.3(2), table-space-text-post = a, table-align-text-post = false]
                S[table-format = 1.3(2), table-space-text-post = a, table-align-text-post = false]
                S[table-format = +1.4(2)]
                S[table-format = 1.3(2)]
                S[table-format = 1.2]}
                        \hline\hline
                        \multicolumn{2}{l}{\multirow{2}{*}{Parameter}}  &\multicolumn{1}{c}{Median}     &\multicolumn{1}{c}{Mean}       &\multicolumn{1}{c}{Exponent}   &\multicolumn{1}{c}{Balance}    &\multicolumn{1}{c}{Seasonal variation}\\
                                &       &\multicolumn{1}{c}{$d'_\text{med}$}    &\multicolumn{1}{c}{$d'_\text{avg}$}    &\multicolumn{1}{c}{$e'$}       &$c'$   &\multicolumn{1}{c}{$\Delta d$ [\%]}\\
                        \hline
                        \multicolumn{2}{l}{Magnetic field [\si{nT}]}    &5.358(25)      &5.705(28)      &-1.662(11)     &\multicolumn{1}{c}{--} &2.8\\
                        \multicolumn{2}{l}{Density [\si{\per\cm\cubed}]}        &5.424(33)      &6.845(47)      &-2.114(20)     &\multicolumn{1}{c}{--} &3.6\\
                        \multicolumn{2}{l}{Temperature [\SI{e4}{\K}]}   &6.357(64)      &10.72(14)      &-1.100(20)     &\multicolumn{1}{c}{--} &1.9\\
                        \hline
                        \multirow{2}{*}{Velocity}       &$W''_1$        &3.707(13)      &3.748(16)      &0.0990(51)    &0.557(45)     &0.17\\
                        \multirow{2}{*}{[\SI{e2}{\km\per\s}]}   &$W''_2$        &5.26(13)       &5.42(11)       &0.0990(51)       &0.557(45)       &0.17\\
                        \cline{2-7}
                                &$W''_\text{II}$        &4.13(13)\tablefootmark{a}      &4.47(11)\tablefootmark{a}      &\multicolumn{1}{c}{--} &\multicolumn{1}{c}{--} &\multicolumn{1}{c}{--}\\
                        \hline
                \end{tabular}
                \tablefoot{
                        \tablefoottext{a}{Velocity median and mean \SI{1}{\au} values for the resulting function. Error estimates derived from the individual fit part errors.}
                }
        \end{table*}

        The velocity balancing parameter $c' = 0.557$ is in good agreement with the results for the SSN dependency (\ref{eq:balance_with_ssn}), because with a mean SSN of 59 during the Helios time period, $c(59) = 0.53$, as can be seen from Fig.~\ref{fig:Vdbl_SSN_ratio_f_plot}.

        The power law lognormal models and the power law double lognormal model for the velocity, which result from the fitting, are plotted in Fig.~\ref{fig:mixed_fit_fixed_4_paper_f_plot} together with their median values. The model’s magnetic field strength is broader around values of \SI{40}{nT} at the lower distance boundary than the data's frequency distribution implies. This behavior is expected because of the applied distance-independent shape approximation. The velocity and temperature models’ upper values generally show a higher abundance than the actual data; see also zoom boxes in Figs.~\ref{fig:histogram_fits_4_a_zoom_paper_pdfplot} and \ref{fig:histogram_fits_V_a_zoom_dbl_paper_pdfplot}. The high-velocity tail that increases with distance arises from using the same exponent for both slow and fast components. This effect is not seen in the data; more specifically, not only the slowest wind but also the fastest wind is expected to converge to more average speeds \citep{Sanchez-Diaz2016}.

        \section{Empirical solar-wind model}
        \label{sec:empirical_solar_wind_model}
        In order to estimate the solar-wind environment for the PSP orbit, we combine the results from the solar-wind frequency distributions’ solar-activity relationships and their distance dependencies derived from the OMNI and Helios data. The result is an empirical solar-wind model for the inner heliosphere which is then extrapolated to the PSP orbit in Sect.~\ref{sec:model_extrapolation_to_psp_orbit}.

\begin{figure}
                \resizebox{\hsize}{!}{\includegraphics{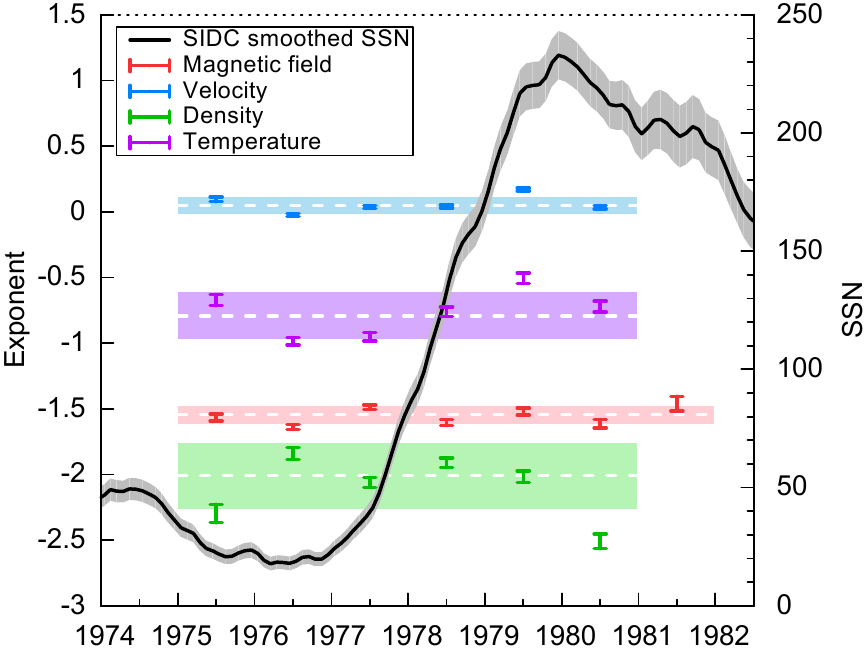}}
                \caption{Helios yearly variation of the solar wind parameter power exponents for the dependence on radial distance together with the SIDC 13-month smoothed monthly SSN. The weighted standard deviations and average values for all years are indicated by the shaded areas. In this plot, the 21~days since Helios launch in the year 1974 are omitted because a distance range of merely \SIrange{0.95}{0.98}{\au} was covered that year.}
                \label{fig:yearly_gradients_c}
        \end{figure}

        This solar-wind model for the radial distance dependence is representative for the time of the Helios observations around the rise of solar cycle~21. The variations of the yearly power law fit exponents from fitting the solar-distance dependency (\ref{eq:power_function}) are shown in Fig.~\ref{fig:yearly_gradients_c} together with the yearly SSN for the time period \numrange{1974}{1982}. It can be seen that during the Helios time period there might be some systematic  variation of the exponents with solar activity -- at least for the velocity and temperature exponents.
        However, for simplicity we assume that the distance scaling laws can be treated as time independent and include the calculated exponents’ yearly variations $\Delta e$, summarized in Table~\ref{tab:mean_median_fit_parameter}, as relative uncertainties.

        Since we neglect possible variations of the distance scaling laws, we combine the frequency distribution’s median solar activity dependency (\ref{eq:median_with_ssn}) derived for \SI{1}{\au} from the OMNI data with the power law exponents (\ref{eq:power_function}) derived from the Helios data:
        \begin{align}
                x_\text{med}(ssn,r) &= \left(a_\text{med} \cdot ssn + b_\text{med}\right) \cdot r^{e'}    \,.     \label{eq:general_sw_model}
        \end{align}
        Thus, implementing the median and mean relations into the lognormal function (\ref{eq:single_lognormal_fit_function}), we obtain the combined model function $W'''(x,ssn,r)$ and for the velocity $W_\text{II}'''(x,ssn,r)$ with the double lognormal function (\ref{eq:double_lognormal_fit_function}). The corresponding median and mean relations for each solar-wind parameter, based on the values resulting from our analyses, are listed below. Their numerical values are the fit parameters from Table~\ref{tab:ssn_fit_parameters} and the exponents from Table~\ref{tab:extrapolation_model_fit_parameters}.

        \begin{itemize}
                \item The magnetic field strength relations, depending on solar activity and solar distance, are:
                \begin{align}
                        B_\text{med}(ssn,r) &= \left(\SI{0.0131}{\nT} \cdot ssn + \SI{4.29}{\nT}\right) \cdot r^{-1.66}     \,,     \label{eq:median_B}\\
                        B_\text{avg}(ssn,r) &= 1.0879 \cdot B_\text{med}(ssn,r)\,.
                \end{align}
                \item The proton velocity relations for the slow and fast components, depending on solar distance, are:
                \begin{align}
                        v_\text{med}^\text{slow}(r) &= \SI{363}{\km\per\s} \cdot r^{0.099} \,,&    &       &v_\text{med}^\text{fast}(r) &= \SI{483}{\km\per\s} \cdot r^{0.099} \,,     \label{eq:median_v}\\
                        v_\text{avg}^\text{slow}(r) &= 1.0101 \cdot v_\text{med}^\text{slow}(r) \,,&    &       &v_\text{avg}^\text{fast}(r) &= 1.023 \cdot v_\text{med}^\text{fast}(r)      \,.
                \end{align}
                The share of both components balanced with solar activity is found to be:
                \begin{align}
                        c(ssn) = -0.00180 \cdot ssn + 0.64      \,.     \label{eq:balance_v}
                \end{align}
                \item The derived relations of the proton density are:
                \begin{align}
                        n_\text{med}(ssn,r) &= \left(\SI{0.0038}{\per\cm\cubed} \cdot ssn + \SI{4.50}{\per\cm\cubed}\right) \cdot r^{-2.11}     \,,     \label{eq:median_n}\\                   n_\text{avg}(ssn,r) &= 1.305 \cdot n_\text{med}(ssn,r)\,.
                \end{align}
                \item The derived proton temperature relations are:
                \begin{align}
                        T_\text{med}(ssn,r) &= (\SI{197}{\K} \cdot ssn + \SI{57300}{\K}) \cdot r^{-1.10} \,,     \label{eq:median_T}\\
                        T_\text{avg}(ssn,r) &= 1.654 \cdot T_\text{med}(ssn,r)\,.
                \end{align}
        \end{itemize}

        These relations average over seasonal variations because they are based on yearly data. The OMNI data are time-shifted to the nose of the Earth’s bow shock; this leads to yearly solar distance variations of \SI{+-1.67}{\%} as it orbits the Sun. The resulting maximal solar-wind parameter variation amplitudes over the year can thus be derived from the derived power law exponents. They are estimated to be smaller than \SI{4}{\percent} as seen in Table~\ref{tab:extrapolation_model_fit_parameters}. \citet{Bruno1986} and \citet{Balogh1999} have pointed out that the solar-wind parameters vary with latitudinal separation from the heliospheric current sheet. Its position in heliographic latitude is highly variable around the solar equator \citep{Schwenn1990} and, furthermore, the Earth’s orbit varies over the course of the year by \SI{+-7.2}{\degree} in latitude. Since this latitudinal separation is highly variable and requires significant effort to calculate for an extended time series, we have ignored this aspect in this analysis.

        \section{Model extrapolation to PSP orbit}
        \label{sec:model_extrapolation_to_psp_orbit}
        To estimate PSP’s solar-wind environment during its mission time for its orbital positions, predictions of the SSN during the mission are incorporated into the empirical solar-wind model, derived in the previous Sections, and extrapolations down to the PSP perihelion region are performed.

\begin{figure}
                \resizebox{\hsize}{!}{\includegraphics{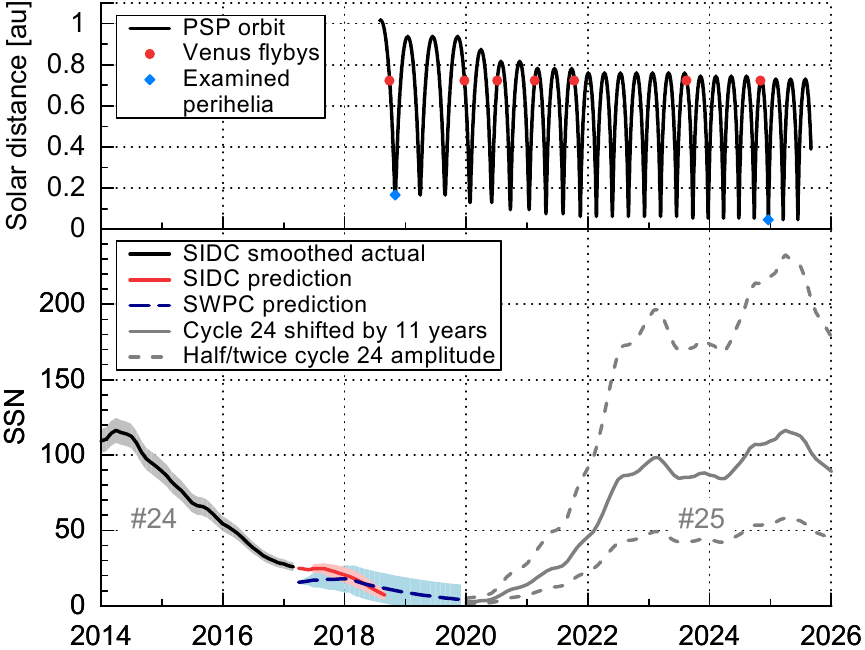}}
                \caption{PSP's solar distance during its mission time (top). Consecutive Venus flybys bring its perihelia nearer to the Sun. Actual and predicted SSN (bottom), that is, SIDC 13-month smoothed monthly actual SSN, SIDC Standard Curves Kalman filter prediction and SWPC prediction with their corresponding expected ranges (shaded areas). The SSN from previous cycle~24, shifted by 11~years, is plotted together with two alternative trends of half and twice its amplitude.}
                \label{fig:SPP_orbit_predicted_SSN_overview_f_plot}
        \end{figure}

        Parker Solar Probe is planned to launch in mid 2018. With its first Venus flyby it will swing into Venus' orbital plane, reaching  a first perihelion with a distance of \SI{0.16}{\au  } just 93~days
after launch, in November 2018. Seven additional Venus flybys allow the perihelion distance to be reduced to a minimum of \SI{9.86}{\Rs}. This distance will be reached with the 22nd perihelion in December 2024 \citep{Fox2015}, as plotted in the top panel of Fig.~\ref{fig:SPP_orbit_predicted_SSN_overview_f_plot}.

\begin{figure*}
                \includegraphics[width=18cm]{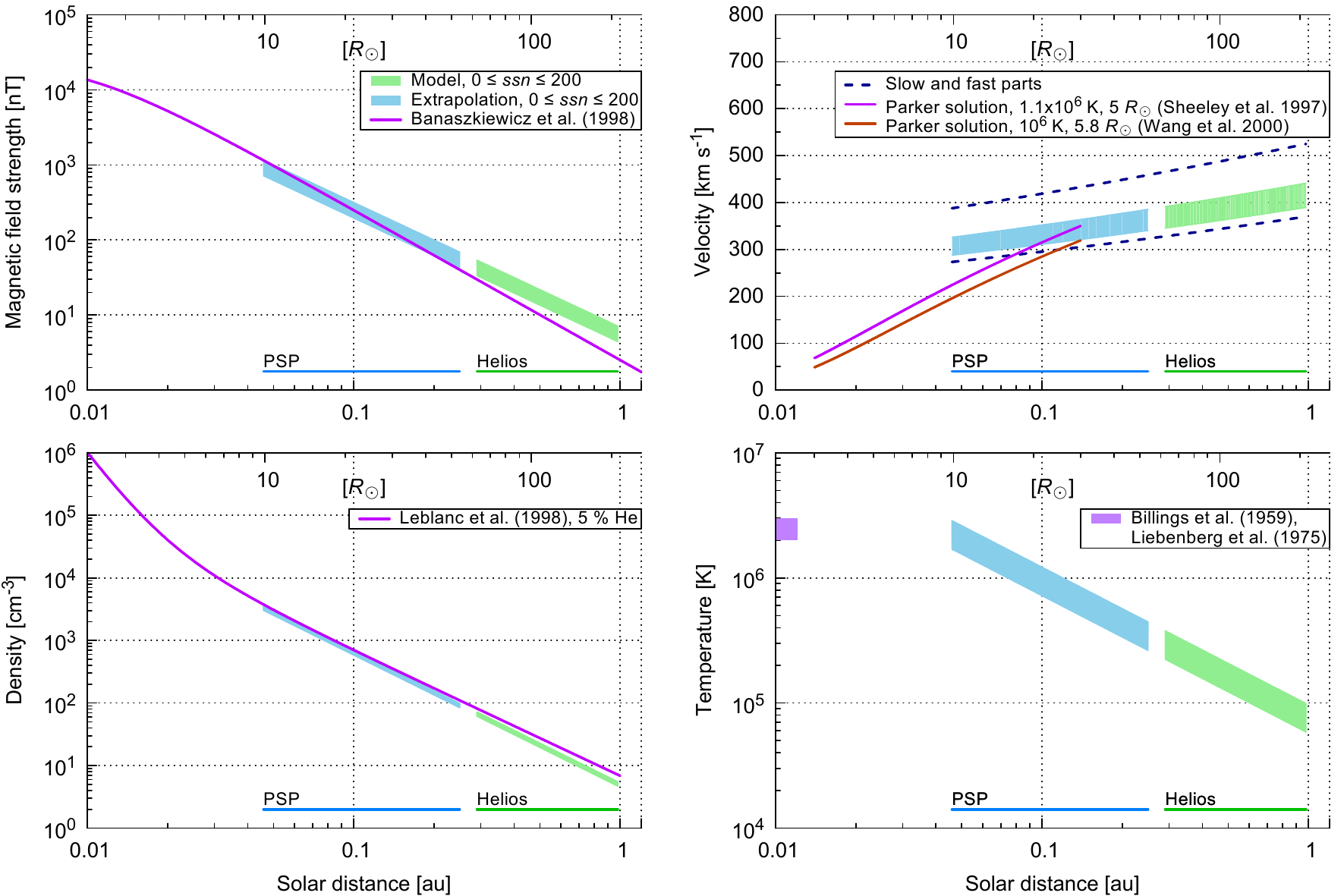}
                \caption{Radial extrapolation of the solar-wind parameters to the PSP orbit region. The median values from the models, obtained from Helios and OMNI measurements, are extrapolated to the PSP region for SSN values between solar minimum and maximum, that is, $0 \le ssn \le 200$. The lower edges of the shaded areas correspond to solar minimum, the upper edges to solar maximum. Also plotted are the radial dependencies of the analytic DQCS magnetic field model for solar minimum from \citet{Banaszkiewicz1998}, the slow wind velocity models from \citet{Sheeley1997} and \citet{Wang2000}, the density model from \citet{Leblanc1998} and the range of temperature measurements from \citet{Billings1959} and \citet{Liebenberg1975}.}
                \label{fig:sw_extrapolation_ssn_f_plot}
        \end{figure*}

        We extrapolate the derived empirical solar-wind models to PSP’s orbital distance range and compare the results with those from the existing models shown in Fig.~\ref{fig:sw_extrapolation_ssn_f_plot}.
        The model and its extrapolation are visualized for the SSN range between solar minimum and maximum ($0 \le ssn \le 200$), indicated by the shaded regions in the Figure.
        The magnetic field strength is found to increase from median values of about \SI{43}{\nT} at \SI{0.25}{\au} to \SI{715}{\nT} at \SI{0.046}{\au} for a SSN of zero. Taking a SSN of 200 increases the values to \SI{69}{\nT} and \SI{1152}{\nT}. Our extrapolation results are slightly flatter than those derived from the analytical magnetic field model by \citet{Banaszkiewicz1998}, who constructed an analytic dipole plus quadrupole plus current sheet (DQCS) model for solar minimum. We note that one cannot easily compare the absolute values of our study with the values obtained by \citet{Banaszkiewicz1998} because the DQCS model assumes solar wind originating from coronal holes at higher heliographic latitudes only, neglecting the slow solar-wind belt.
        We suggest that the difference in slope is due to the previously mentioned (Sect.~\ref{sec:solar_distance_dependency}) changing shape of the frequency distribution with heliocentric distance, which for smaller distances deviates more from the model’s lognormal distribution.
        The average velocity is found to decrease from \SI{340}{\km\per\s} at \SI{0.25}{\au} to about \SI{290}{\km\per\s} at \SI{0.046}{\au} 3~years after a SSN of zero occurred, whereas using a SSN of 200 it decreases from \SI{390}{\km\per\s} to \SI{330}{\km\per\s}. Comparing the results with the measurements by \citet{Sheeley1997} and \citet{Wang2000} shows an overestimation in our extrapolated slow solar-wind velocity values for distances below approximately \SI{20}{\Rs}. They used LASCO coronagraph observations to track moving coronal features (blobs) in the distance range \SIrange{2}{30}{\Rs} to determine speed profiles and sources of the slow solar wind and they derived temperature and sonic point values for slow solar wind with the isothermal expansion model from \citet{Parker1958}. Therefore, it generally can be expected that PSP will encounter a slower solar-wind environment close to the Sun than our model estimates. Thus PSP will measure solar-wind acceleration processes \citep{McComas2008}, maybe even still at \SI{30}{\Rs} as the study by \citet{Sheeley1997} suggests.
        The proton density increases from about \SI{84}{\per\cm\cubed} at \SI{0.25}{\au} to about \SI{3018}{\per\cm\cubed} at \SI{0.046}{\au} 6~years after a SSN of zero occurred. Being almost independent of the SSN, the values for a SSN of 200 are only \SI{17}{\%} larger. The results are in good agreement with those of \citet{Leblanc1998}, who derived an electron density model from type~III radio burst observations. Their model shows that the density distance dependency scales with $r^{-2}$ and steepens just below \SI{10}{\Rs} with $r^{-6}$. We assumed a solar-wind helium abundance of \SI{5}{\%} to convert these electron densities to proton densities.
        The extrapolated proton temperature increases from about \SI{260000}{\K} at \SI{0.25}{\au} to about \SI{1690000}{\K} at \SI{0.046}{\au} 3~years after a SSN of zero occurred and from \SI{440000}{\K} to \SI{2860000}{\K} for a SSN of 200. Knowing that near-Sun coronal temperatures are in the range of \SIrange{2}{3}{\mega\K} \citep{Billings1959,Liebenberg1975}, the model overestimates the extrapolated temperatures at the PSP perihelion distance.

        Aside from the solar distance, the derived solar-wind parameter models depend on the SSN. Short-term predictions of the SSN can be used for the solar-wind predictions of PSP's early perihelia and also for refining the solar-wind predictions during PSP's mission. Several sources are available for SSN short-term predictions. The SIDC provides 12-month SSN forecasts\footnote{\url{http://sidc.be/silso/forecasts}} obtained from different methods (e.g., Kalman filter Standard Curve method). The SSN prediction of NOAA's Space Weather Prediction Center (SWPC) for the time period until the end of 2019 follows  a consensus of the Solar~Cycle~24 Prediction~Panel\footnote{\url{http://www.swpc.noaa.gov/products/solar-cycle-progression}}.
        The SSN for PSP's first perihelion will be small -- certainly below 20 -- whereas the SSN during the closest perihelia, which will commence at the end of 2024 at the likely maximum phase of cycle~25, cannot be predicted at this time. However, \citet{Hathaway2016} found indications that the next solar cycle will be similar in size to the current cycle~24. Therefore we simply assume a pattern similar to the last cycle for the prediction of the next solar cycle and thus shift the last cycle by 11~years. Additionally, we consider as possible alternatives SSN patterns of half and twice its amplitude as shown in the bottom panel of Fig.~\ref{fig:SPP_orbit_predicted_SSN_overview_f_plot}.

\begin{figure}
                \resizebox{\hsize}{!}{\includegraphics{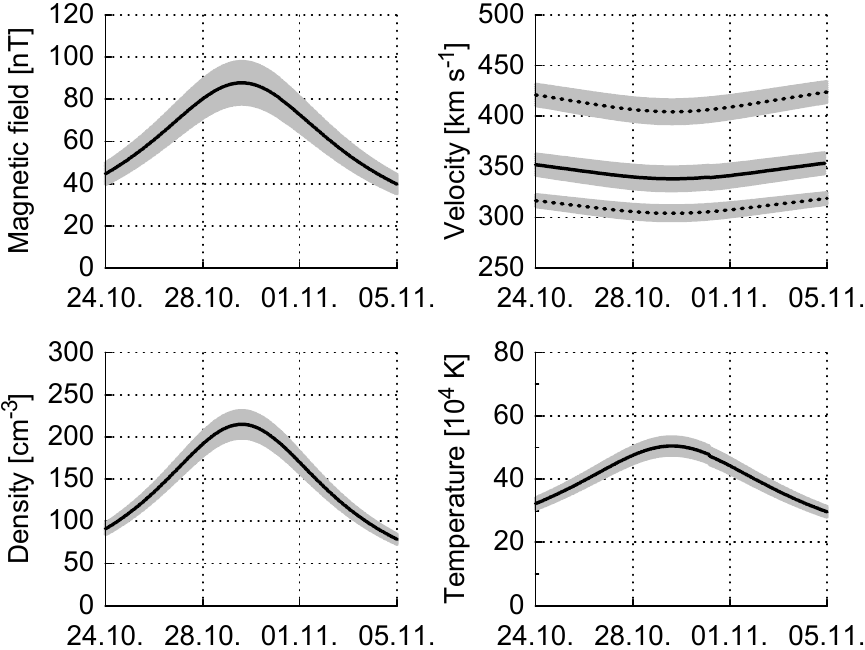}}
                \caption{Estimated solar-wind parameter medians (black lines) and their error bands (gray) during 12~days in 2018 with PSP's first perihelion at about \SI{0.16}{\au}. For the velocity the combined median is calculated and also the SSN-independent slow and fast parts are plotted (dotted lines).}
                \label{fig:SPP_perihelia_prediction_f_plot}
        \end{figure}

\begin{figure}
                \resizebox{\hsize}{!}{\includegraphics{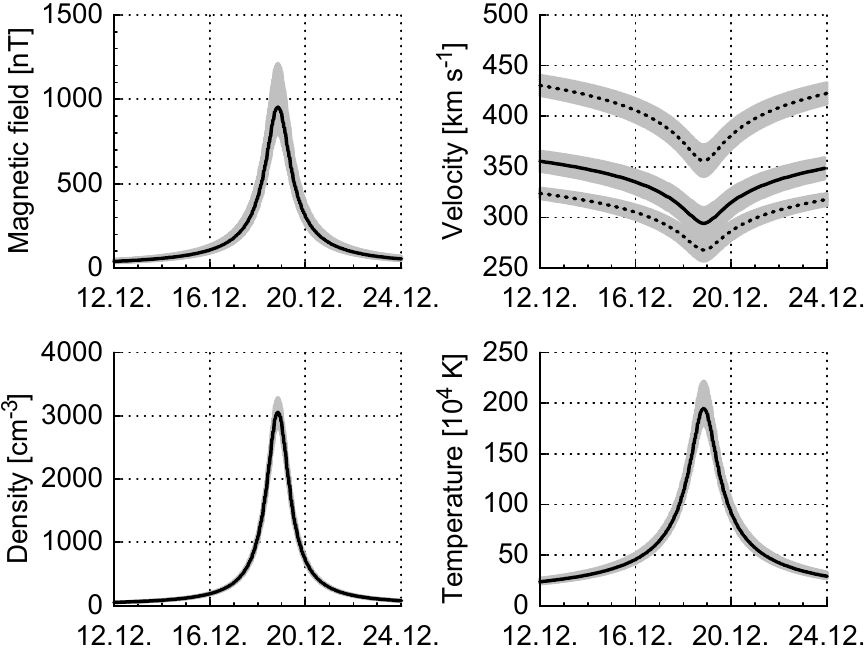}}
                \caption{Estimated solar-wind parameter medians (black lines) and their error bands (gray) during 12~days in 2024 with PSP's 22nd (the first closest) perihelion at \SI{0.0459}{\au}. For the velocity the combined median is calculated and also the SSN-independent slow and fast parts are plotted (dotted lines).}
                \label{fig:SPP_perihelia_prediction_nearest_f_plot}
        \end{figure}

        Implementing the SSN predictions for the PSP mission time and the orbital trajectory data, we can infer which solar-wind parameter magnitudes can be expected. Figures~\ref{fig:SPP_perihelia_prediction_f_plot} and \ref{fig:SPP_perihelia_prediction_nearest_f_plot} show the median values (\ref{eq:median_with_ssn}) of the considered different solar-wind parameters for 12-day periods, comprising the first perihelion in Novemver 2018 and the first closest perihelion in December 2024.
        In the beginning of the mission median values of about \SI{87}{\nT}, \SI{340}{\km\per\s}, \SI{214}{\per\cm\cubed} and \SI{503000}{\K} are estimated to be measured at \SI{0.16}{\au}, increasing to about \SI{943}{\nT}, \SI{290}{\km\per\s}, \SI{2951}{\per\cm\cubed} and \SI{1930000}{\K} during the first closest approach at \SI{0.046}{\au}. Monthly SSNs -- shifted by the time lags specific to the solar-wind parameters -- are used in the calculation of the solar-wind predictions. These SSNs are either actual smoothed values from the SIDC with their reported standard deviations, short-term predictions from the SWPC with their expected ranges, or actual smoothed values from the SIDC shifted by 11~years with half/twice their values as uncertainties. The error bands given in both Figures, calculated from error propagation, include these SSN ranges and the derived fit parameter errors.

\begin{figure}
                \resizebox{\hsize}{!}{\includegraphics{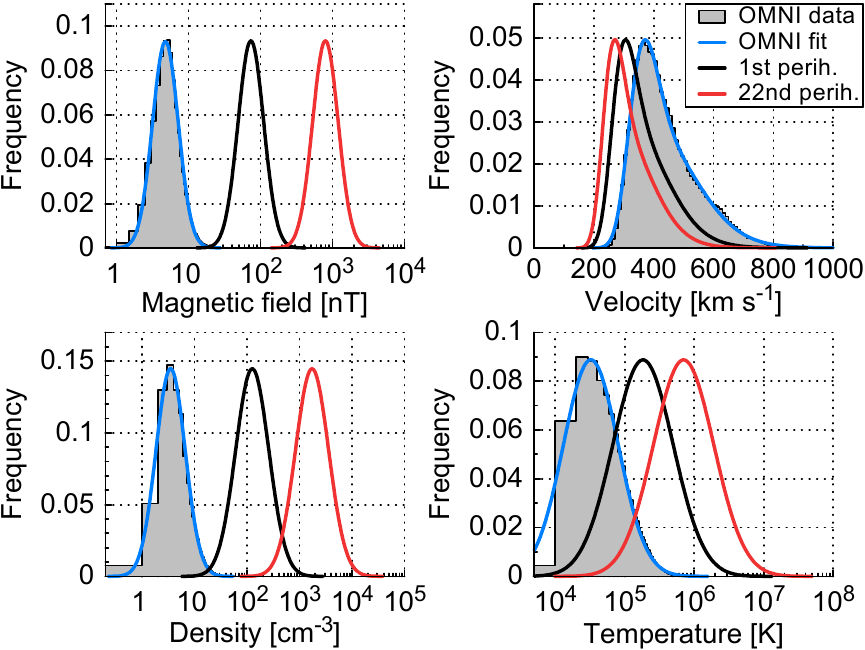}}
                \caption{Frequency distributions of the four solar-wind parameters (same as in Figs.~\ref{fig:histogram_fits_4_a_zoom_paper_pdfplot} and \ref{fig:histogram_fits_V_a_zoom_dbl_paper_pdfplot}) and those estimated with the solar-wind model for PSP's 1st and 22nd (first closest) perihelion. In these Figures the frequencies of both extrapolated curves are scaled for visibility to the same height as the \SI{1}{\au} distribution.}
                \label{fig:SPP_sw_distributions_b}
        \end{figure}

        Finally the estimated solar-wind environment can be derived from the function $W'''(x,ssn,r)$. The estimated frequency distributions of the four solar-wind parameters at PSP's 1st and 22nd (first closest) perihelion are plotted in Fig.~\ref{fig:SPP_sw_distributions_b}.
        Again, we point out that the velocity and temperature distributions for the 22nd perihelion are only upper limits and the actual values to be encountered by PSP are expected to be smaller.

        \section{Discussion and summary}
        \label{sec:discussion_and_summary}
        The scientific objective of this study, being part of the CGAUSS project -- the German contribution to the WISPR instrument -- is to model the solar-wind environment for the PSP mission to be launched mid 2018. For this purpose we started the development of the empirical solar-wind environment model for the near-ecliptic PSP orbit. We derived lognormal representations of the in situ near-Earth solar-wind data collected in the OMNI database, using the frequency distributions of the key solar-wind parameters, magnetic field strength, proton velocity, density, and temperature. Throughout the different analyses in our study, the velocity's frequency distribution is treated as a composition of a slow and a fast wind distribution. Each velocity part is fitted with a lognormal function, which allows for the overlap of both velocity ranges. The OMNI multi-spacecraft solar-wind data is intercalibrated and covers almost five solar cycles. It thus represents solar wind gathered at different phases of solar activity in the ecliptic plane. In the next step we investigated the yearly variation of the solar-wind distribution functions along with the SSN over 53~years and derived linear dependencies of the solar-wind parameters with the SSN. The radial dependencies of the solar-wind distribution functions were then analyzed, using Helios~1 and 2 data for the distance range \SIrange{0.29}{0.98}{\au} in bins of \SI{0.01}{\au}, deriving power law fit functions that were used to scale the previously calculated SSN-dependent \SI{1}{\au} distribution fit functions to the PSP orbit, taking into account SSN predictions for the years 2018--2025, encompassing the prime mission up to the closest approach of \SI{9.86}{\Rs}. The reason for performing the analysis this way is based on the fact that the OMNI solar-wind database is much larger than the Helios database.

        For determining solar-activity- and solar-distance-dependent relations for the median and mean solar-wind values, we could have used the simpler approach of combining the radial dependence of averaged Helios data with averaged \SI{1}{\au} OMNI data scaled with the SSN. It is expected that the results of a simpler analysis would have similar distance scaling results, as can be inferred from the exponents in Tables~\ref{tab:mean_median_fit_parameter} and \ref{tab:extrapolation_model_fit_parameters}. However, in our study we are not only interested in averages but rather in bulk distributions, that is, the whole range of values that might occur. For the determination of the frequency distributions the use of the more complex fit model is important, because the distance between median and mean values determines the width of the lognormal distributions.

        It is clear that the calculated distribution functions only represent first-order estimates of the real solar wind to be encountered by PSP. The solar-wind environment to be encountered will depend at times of PSP on the structure of the solar corona and underlying photospheric magnetic field and on the evolution and interaction of individual solar-wind streams and superimposed CMEs and shocks. However, the derived results are in good agreement with existing studies about near-Sun solar-wind magnetic field strengths and densities as shown in Sect.~\ref{sec:model_extrapolation_to_psp_orbit}. The extrapolation results of the velocity and the temperature differ from the direct measurements seen in existing studies. This suggests that below about \SI{20}{\Rs} PSP may dive into the region where the acceleration and heating of the solar wind is expected to occur (see Fig.~\ref{fig:sw_extrapolation_ssn_f_plot}). The near-Sun solar-wind velocity at PSP perihelion is also expected to be slower than our model estimates, because the solar wind is assumed to be accelerated up to the height of the Alfvénic critical surface, which is predicted to lie on average around \SI{17}{\Rs} \citep[e.g.,][]{Sittler1999,Exarhos2000}, scaling with solar activity within a range of between \SI{15}{\Rs} at solar minimum and \SI{30}{\Rs} at solar maximum \citep{Katsikas2010,Goelzer2014}.

        We have not specifically investigated the occurrences of extreme solar-wind parameters caused by CMEs or enhanced values due to stream interaction or co-rotating interaction regions. The Helios solar-wind measurements plotted over radial distance in Fig.~\ref{fig:radial_fit_4_thesis_light_b_skip} show several extreme values far above the usual solar-wind velocities, which are associated with individual CMEs. The results by \citet{Sachdeva2017} indicate that due to solar-wind drag, the speeds of fast CMEs will commonly slow down substantially from early distances of a few solar radii. Therefore, it is expected that PSP will encounter CMEs with much higher speeds than those observed during the Helios mission. Also, the magnetic field, density and temperature values are expected to be much larger than in the average solar wind in individual fast-shock-associated CME events. PSP will thus also substantially improve our understanding of the near-Sun evolution of CMEs and their expansion with radial distance.

        With the resulting CGAUSS empirical solar-wind model for PSP, the following main results for the bulk solar-wind parameters and estimations for their median values at PSP’s first perihelion in 2018 at a solar distance of \SI{0.16}{\au} and at PSP’s closest perihelia beginning in 2024 at \SI{0.046}{\au} (\SI{9.86}{\Rs}) are obtained:
        \begin{itemize}
                \item The dependency of the magnetic field strength on solar activity and radial distance appears to be valid above \SI{20}{\Rs}, however near PSP's closest perihelia, the actual values might be found to be slightly higher.
                \item The estimated magnetic field strength median values obtained from relation (\ref{eq:median_B}) for PSP's 1st and 22nd perihelion are \SI{87}{\nT} and \SI{943}{\nT}.
                \item The radial dependencies of the proton velocity median values for slow and fast solar wind (\ref{eq:median_v}) appear to be valid above about \SI{20}{\Rs} solar distance; below they overestimate the actual solar wind velocities obtained from remote measurements. The share of their frequency distributions to the overall solar-wind velocity distribution (\ref{eq:double_lognormal_fit_function}) depends on solar activity with their balance relation (\ref{eq:balance_v}). Thus, at solar minimum, with a SSN of around zero, the slow-wind component contributes about \SI{64}{\%} and drops to \SI{28}{\%} during solar maximum conditions with a SSN around 200.
                \item The calculated median velocity values for PSP's 1st and 22nd perihelion are \SI{340}{\km\per\s} and \SI{290}{\km\per\s}.
                \item The proton density relation appears to be valid throughout the full PSP orbital distance range, even down to about \SI{8}{\Rs}.
                \item The estimated density median values obtained from relation (\ref{eq:median_n}) for PSP's 1st and 22nd perihelion are \SI{214}{\per\cm\cubed} and \SI{2951}{\per\cm\cubed}.
                \item The derived correlation function for the proton temperature appears to provide overly high temperature values around PSP’s closest perihelion in comparison to coronal measurements.
                \item The estimated temperature median values obtained from relation (\ref{eq:median_T}) for PSP's 1st and 22nd perihelion are \SI{503000}{\K} and \SI{1930000}{\K}.
        \end{itemize}

        The results of the modeled solar-wind environment will be useful to help optimize the WISPR and in situ instrument science plannings and PSP mission operations. This also applies for the Heliospheric Imager (SoloHI) \citep{Howard2013} and the in situ instruments on board the Solar Orbiter spacecraft.

        \begin{acknowledgements}
                The authors acknowledge support of the Coronagraphic German and US SolarProbePlus Survey (CGAUSS) project for WISPR by the German Aerospace Center (DLR) under grant 50~OL~1601 as national contribution to the Parker Solar Probe mission. The authors thank the Helios and OMNI PIs/teams for creating and making available the solar wind in situ data. The Helios and the OMNI data are supplied by the NASA Space Science Data Coordinated Archive and the Space Physics Data Facility at NASA's Goddard Space Flight Center. Additional thanks for maintaining and providing the international sunspot number series goes to the World Data Center -- Sunspot Index and Long-term Solar Observations at the Solar Influences Data Analysis Center, Royal Observatory of Belgium. The PSP SPICE kernel was kindly provided by Angelos~Vourlidas. The authors thank the referee for the careful review of this manuscript and helpful comments and suggestions.
        \end{acknowledgements}

        \bibliographystyle{aa}

\end{document}